\documentclass[aps,prb,twocolumn,groupedaddress,showpacs,showkeys]{revtex4}

\usepackage{graphicx,amsmath,amssymb,amsfonts,latexsym,color,dcolumn,bm,subfigure,ulem}

\begin{document}

\newcommand{\be}{\begin{equation}}
\newcommand{\ee}{\end{equation}}
\newcommand{\rd}[1]{\textcolor{red}{#1}}
\newcommand{\bl}[1]{\textcolor{blue}{#1}}
\newcommand{\ma}[1]{\textcolor{magenta}{#1}}
	
	\title{Isoelectronic determination of the thermal Casimir force}

	\author{G. Bimonte}
	\affiliation{Dipartimento  di  Fisica,  Universit\`a  di  Napoli  Federico  II, Complesso  Universitario  MSA,  Via  Cintia,  I-80126  Napoli,  Italy  and INFN  Sezione  di  Napoli,  I-80126  Napoli,  Italy}
	\author{D. L\'opez}
	\affiliation{Center for Nanoscale Materials, \\Argonne National Laboratories, Argonne, Illinois 60439, USA}
	\author{R. S. Decca}
	\email{rdecca@iupui.edu}
	\affiliation{Department of Physics, \\Indiana University-Purdue University Indianapolis, Indianapolis, Indiana 46202, USA}

	\date{ \today}
	
	\begin{abstract}
		Differential force measurements between spheres coated with either nickel or gold and rotating disks with  periodic distributions of nickel and gold are reported. The rotating samples  are covered by a thin layer of titanium  and a layer of gold. While  titanium is used for fabrication purposes, the gold layer (nominal thicknesses of 21, 37, 47 and 87 nm) provides an isoelectronic environment, and is used to nullify the electrostatic contribution but allow the passage of long wavelength Casimir photons. A direct comparison between the experimental results and predictions from Drude and plasma models for the electrical permittivity  is carried out. In the models the magnetic permeability of nickel is allowed to change to investigate its effects. Possible sources of errors, both in the experimental and theoretical sides, are taken into account. It is found that a Drude response with magnetic properties of nickel taken into account is unequivocally ruled out. The full analysis of the data indicates that a  dielectric plasma response with magnetic properties of Ni included shows  good agreement with the data. Neither a Drude nor a plasma dielectric response provide a satisfactory description if the magnetic properties of nickel are disregarded. 
	\end{abstract}
	
	
\maketitle

\section{Introduction}
Dispersive forces induced by vacuum fluctuations are ubiquitous in nature. The study of these forces between electrically neutral yet polarizable materials is important in disciplines ranging from chemistry to nanotechnology. Historically called van der Waals interaction when the separation of the bodies is small (non-retarded regime)\cite{vdW} and Casimir forces at larger separations (retarded regimes)\cite{Casimir} it is clear these forces have a common origin. While for many years a quantitative experimental confirmation of the Casimir interaction remained elusive, in the last couple of decades the precision and complexity of experimental determinations of the Casimir interaction has yielded a plethora of results. Starting with the pioneering work of S. Lamoreaux\cite{RSDLamoreaux1997} and U. Mohideen,\cite{RSDMohideen1998} measurements  using different geometries\cite{RSDEderth2000,RSDBressi2001,Brown2005}, using micromechanical oscillators \cite{RSDChan2001a,RSDTang2012,RSDChan2013}, between different materials \cite{RSDDecca2003,RSDIanuzzi2004,RSDMohideen2005,RSDLamoreaux2009,RSDMunday2009,deMan,RSDPalasantzas2010,RSDMohideen2012}, and at low temperatures\cite{RSDChan2013,RSDDecca2009,RSDMohideen2013b} (to mention some examples) followed. For recent reviews see [\onlinecite{reviewCasimir,book2,capasso}].

It was not, however,  until precise enough measurements of the Casimir force allowed for a quantitative comparison with theoretical models that significant issues appeared in the understanding of vacuum fluctuations in the presence of dielectric boundaries.\cite{reviewCasimir,book2,decca4}
Lifshitz showed that, in addition to quantum fluctuations, thermal fluctuations of the electromagnetic fields provide an extra contribution  to  the  Casimir  force,  called 	thermal Casimir force.\cite{Lifshitz} A different thermal contribution to the Casimir force is obtained depending on whether the complex dielectric function of the conductor is extrapolated to zero frequency on the basis of the  Drude  model or  instead  by  the  dissipationless  plasma  model.\cite{reviewCasimir, book2,decca4} Experimentally, the situation is not clear. The most precise measurements  of the Casimir interaction done at submicron separations between two Au bodies \cite{DeccaEJP2007} are very well described by the plasma model, excluding the Drude model with very high confidence. The same holds true for measurements done in Mohideen's group.\cite{MohideenPRA2012} Their judicious selection of materials showed that the experimental data always agreed with the plasma model. This is the case even though the predicted relative strength of the interaction from the two models (plasma or Drude) changes from larger, equivalent or smaller depending on the selection of materials used. On the other hand, experiments performed at separations in the $\sim~1-10~\mu$m range (more relevant for thermal effects since the thermal wavelength $\lambda_{\rm T}= \hbar c/(k_{\rm B}T) \simeq 8~\mu$m at $T = 300$~K) yielded a better agreement with the Drude prescription once a fit of the electrostatic background was subtracted.\cite{LamoreauxNat2011}
	
One of us recently introduced an approach \cite{BimontePRL2014} towards elucidating the role of dissipation in the thermal contribution to the Casimir force. If one of the interacting bodies is covered by a layer of thickness $ t > \delta $ of a conductor with skin depth $\delta$, the reflectivity of the compound sample for frequencies $\omega > \omega_{\rm c} \simeq c/2 t$ is governed by the top layer. On the other hand, the reflectivity of the compound sample at $\omega < \omega_{\rm c}$ carries information of the underlying structure. In consequence, for measurements performed at room temperature where the thermal component of the free energy per unit area is dominated by the zero Matsubara term, the thin layer of thickness $t$ effectively enhances the difference between the Drude and plasma models when used to extrapolate towards zero frequency. In this paper  these ideas developed in Ref.~[\onlinecite{BimontePRL2014}] are  experimentally implemented using an  approach similar to the one used in Ref.~[\onlinecite{DeccaPRL2014}]. A rotating sample made out of alternating Au-Ni sectors capped by a thin layer of Au allows to fully utilize the high force sensitivity provided by the large mechanical quality of microelectromechanical torsional oscillators. The setup directly yields the low frequency component of the difference between the interaction of a common probe (a metal coated sphere) with the different regions (Au or Ni) of the Au-capped rotating disk. Furthermore, since the only dependence of the signal at the spatial frequency provided by the Au/Ni sectors arises from the difference in their Casimir interactions with a probing metal-coated sphere, all background induced systematics (including the effect of ``patch'' potentials\cite{patches}) are significantly reduced by means of a lock-in detection scheme.  

The plan of the paper is as follows.  Section~\ref{exp} describes the experimental setup. In Section~\ref{data} the data acquisition process, the influence of systematic errors and their subtraction procedure, and the experimental data  obtained are shown. In Section~\ref{calc} the details of the calculation of the Casimir force in the experimental configuration are presented. In Section~\ref{comp} a quantitative comparison between experiments and theory is  presented. Section~
\ref{concl} presents the conclusions. Appendices show details of the calculations that can be omitted in a first reading of the paper.

\section{Experimental setup: Basics for the measurements\label{exp}}

\begin{figure}[htbp]
	\centerline{\includegraphics[width=9cm]{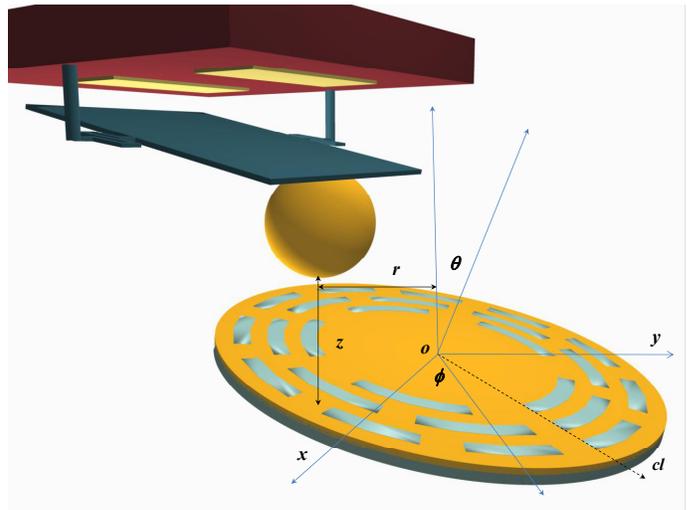}}
	\caption{(Color online) Schematic of the experimental setup. Three regions with $n = 5, 8, 11$ Au-Ni sectors are shown. The actual sample has $n = 50, 75, \cdots, 300$. The region with $n = 50$ has inner and outer  radii of $^{50}r_{\rm i}=4.00$~mm and $^{50}r_{\rm o}=4.15$~mm. A gap of 200~$\mu$m follows. All gaps have the same radial extent and all sectors have $^nr_{\rm o}- ^nr_{\rm i}=150~\mu$m. The $\{x,y\}$ plane defines the plane of rotation of the spindle, selected to be parallel to the MTO's substrate. $cl$ is the line where all regions with different $n$ have a Au-Ni interface. $\theta$ is the change in the instantaneous axis of rotation, $\phi =\omega t$ is the angle or rotation. The distance $z$ is determined from the vertex of the metal-covered sphere to the rotating sample. $r$ is the distance from the sphere's vertex to the center $o$ of the rotating sample. Displacements $\Delta r$ between $o$ and the axis of rotation and the  Ti-Au film covering the rotating sample are not shown.}
	\label{schem}
\end{figure}

The basis of the apparatus and the measurement technique can be seen in Fig.~\ref{schem}. At the heart of the experiment, which allows for a very high force sensitivity and reproducibility, is a large quality factor $Q$ micro mechanical torsional oscillator (MTO). The sample made of $n$ alternating Au and Ni sectors is forced to rotate at an angular frequency 

\be
\omega=2\pi \frac{f_{\rm r}}{n} 
\label{omega}
\ee

\noindent where $f_{\rm r}$ is the operating resonant frequency of the MTO. The first harmonic of the force associated with the angular distribution of the sample will be then naturally selected by the MTO. All other harmonics of the periodic force and all forces with different angular dependences are outside of the resonance peak of the MTO and consequently ``filtered'' by the sharp $\Delta f \simeq 40$~mHz resonance peak of the oscillator.

The sphere-MTO system is mounted onto a piezo-driven 3 axis computer controlled flex system (MadCity Labs) with a stability better than 0.1~nm over 10 hours on all three axis. The piezo stage is in turn mounted on a stepper-motor driven 5 axis stage (Newport). After the initial alignment is achieved the five stages are locked into place to preclude drift. Extreme care is taken to ensure that all non-metallic parts are covered with Au-coated mylar or Au-coated Al-foil. Furthermore, since it has been observed that Al surfaces closer than 5~cm to the sample produce a drift of electrostatic nature, all Al surfaces on the stepper-motor driven stage and the motors themselves were covered with Au-coated mylar. The mechanical arm between the rotating sample and the MTO is close to 10~cm. While the temperature in the chamber is controlled to better than 0.1~K drifts of about 10~nm/hr are observed. The relative drift between the MTO and the rotating sample is  monitored by continuously measuring the capacitance between an L-shaped piece attached to the MTO holder and two orthogonal plates attached to the base of the vacuum chamber.\cite{Kolb}  A two-color interferometer (see sub-section~\ref{MTO}) is used to monitor the $z$ axis separation. Minimum detectable changes $\sim 0.1$~nm along all three axis are counteracted by supplying the appropriate signal to the piezo stage. The assembly with motion stages and MTO holder  is rigidly mounted into the  vacuum chamber, differently from previous measurements of the Casimir force done by this group. \cite{DeccaEJP2007} As a precautionary measure, all metallic surfaces facing the experimental setup are Au-coated. The whole vacuum chamber is mounted into an actively controlled air-damping table. The table and all connections, both electrical and mechanical,  are isolated from vibration sources  by sand boxes. The combination of vibration isolation systems yielded peak-to-peak vibrations with $z_{\rm pp} < 0.02$~nm
(the detection limit in the accelerometer) for frequencies above 10~Hz. More importantly, external vibrations are not sensed by the MTO, see sub-section~\ref{MTO}. The high quality factor in the oscillator is achieved by pumping the system to $P\leq 10^{-6}$~torr (maintained during each run) by a combination of mechanical, turbomolecular and chemical pumps. 

The air-bearing spindle is produced by KLA-Tencor. The thin air-layer between the rotor and its encasing makes the system very compliant. On the other hand, the large air flow needed to operate the spindle required the design and construction of a special seal. Towards this goal a groove (9.65~mm inner diameter, 1~mm wide and 200~$\mu$m deep)  was machined on the top of the lower plate of the vacuum chamber and filled with high molecular weight diffusion pump oil. A skirt with a matching protrusion is rigidly attached to the rotor, with the protrusion into the groove. In order to diminish oil contamination inside the chamber a water cooled waffle is attached to the top of the plate inside the vacuum chamber. The waffle has a $\sim 40-50~\mu$m clearance with the top of the rotating skirt. 

\subsection{Sample preparation and characterization\label{char}}

In the experiment sapphire spheres with  nominal radii $r_{\rm s} = 150~\mu$m were used. The two spheres used were covered with a thermally evaporated $t_{\rm Cr} \sim 10$~nm layer of Cr to improve adhesion followed by a thermally evaporated $t_{\rm met} \sim 250$~nm film of either Au or Ni. The Au-covered sphere had a SEM determined (149.3 $\pm$ 0.2)~$\mu$m-radius while the Ni-coated one had a (150.8 $\pm$ 0.2)~$\mu$m-radius. The spherical surfaces were characterized by atomic force microscopy (AFM) images. The rms roughness were $t^{\rm Au}_{\rm rms} =0.34$~nm and  $t^{\rm Ni}_{\rm rms} =0.34$~nm for the Au- and Ni-coated spheres respectively. 

The rotating sample was fabricated by e-beam evaporating a $d_{\rm Ti} = (10 \pm 1)$~nm thick layer of Ti on a 1 inch diameter $100~\mu$m thick [100] oriented Si wafer. A $d_{\rm tm} = (2.10 \pm 0.02)~\mu$m thick layer of Au was deposited by thermal evaporation on top of the Ti covered Si wafer. Using conventional photolithography, a photoresist structure  consisting of concentric sectors (see Fig.~\ref{schem}) was defined on the Au. The Au not covered by the photoresist was removed down to the Ti layer with a 4g:2g:10ml 	KI:I$_2$:H$_2$O at 70~$^{\rm o}$C (etching rate $\sim$~250~nm/min). After removing the photoresist a thick ($\sim 3~\mu$m)  Ni film was thermally evaporated and the structure mechanically polished without exposing the Au structure. The sample was glued with NOA61 UV curing cement to a BK7 Schott glass flat with the original Si wafer exposed. The wafer was etched away using KOH, and then a gold layer of thickness $t$ was deposited by thermal evaporation. Four different samples with $t$~=~21, 37, 47 and 84~nm where measured. In all cases the error in $t$ was estimated to be $\delta t$~=~1~nm. It was observed that KOH actually attacked the Ti layer. Measurements performed on similar films set the thickness reduction on the Ti at $\sim$~2~nm. Exposed Au surfaces were characterized by white light interferometry (WLI) and AFM. Both techniques showed an optical quality film with no memory  of the underlying structure. The  1024 $\times$ 1024 pixel$^2$ AFM images obtained over different  $10 \times 10~\mu{\rm m}^2$ regions yielded position independent $\sim$~40~nm peak-to-peak roughness.  In each image there are a two to five isolated spikes $\sim 30$~nm tall and about 100~nm across. Excluding these spikes, the sample has a rms roughness of 0.5 nm. The disk was mounted on the air bearing spindle. It was optically verified that the center of the disk and the axis of rotation of the spindle coincided to better than $\Delta r \sim ~10~\mu$m. The flatness and alignment of the sample were checked {\it in-situ} using a fiber interferometer (response time 10 ms). It was found that the surface of the sample was perpendicular to the axis of rotation to better than $z_{\rm o}$~=~20~nm at $^{300}r$ when rotating the disk at $\omega = 2\pi$~rad/s. 

Significant precautions were taken to ensure the Ni samples (either in the rotating sample or the spheres) did not show a net magnetization. All procedures and measurements were done in non-magnetic environments with stray magnetic fields reduced to $\delta H < 10^{-3}$~Gauss, which was accomplished by using $\mu$-metal shielding. The magnetization of Ni samples prepared similarly to the ones in the rotating sample or the sapphire sphere were measured using SQUID magnetometry, and their magnetization is consistent with zero to within the 10$^{-11}$~emu resolution of the apparatus. 

\begin{figure}[htbp]
	\centerline{\includegraphics[width=9cm]{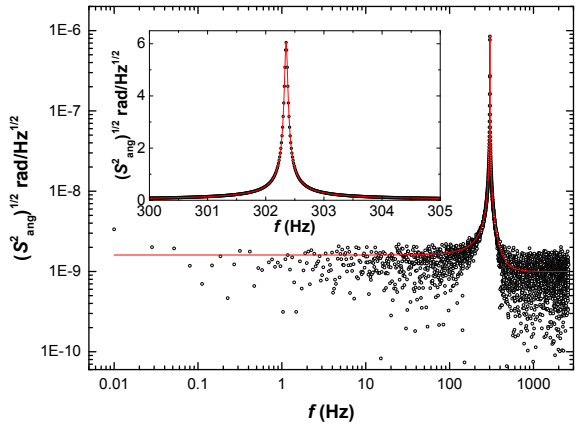}}
	\vspace{-0.3cm}
	\caption{(Color online) Free standing frequency response of the oscillator with the Au-coated sphere glued to it. The square root of the power spectral density shows the limits of the electronic detection circuitry. The inset shows an expanded view of the resonance with an average of 100 different spectra. The red solid line is a fit using Eq.~(\ref{power}) with a detection flat spectral density noise of $1.2 \times 10^{-9}~{\rm rad}/\sqrt{\rm Hz}$. Points below 0.03~Hz where $1/f$-noise is measurable have been excluded for the fit.}
	\label{osc}
\end{figure}

\subsection{Oscillators\label{MTO}}

The MTOs are similar to the ones used in previous experiments. \cite{RSDDecca2003,decca4,DeccaEJP2007} Differently from previous measurements and as schematically shown in Fig.~\ref{schem} the metal coated spheres were glued close to the edge  of the $500~ \times~500~ \mu{\rm m}^2$  plate of the oscillator. Gluing the Au- or Ni-covered  spheres  at distances $b_{\rm Au} = (235 \pm 4)~ \mu$m or $b_{\rm Ni} = (233 \pm 4)~\mu$m from the axis of rotation reduced the MTO's natural frequency of oscillation from $f_{\rm o} \simeq 700$~Hz to $f_{\rm r} = (306.89 \pm 0.05)$~Hz or $f_{\rm r} = (302.57 \pm 0.05)$~Hz respectively. The respective $Q$s were reduced from $\sim 9000$ to $Q_{\rm Au} =4823$ and $Q_{\rm Ni} =5337$.

The power spectral density $S^2_{\rm ang}(f)$ of the oscillator is shown in Fig.~\ref{osc}. For a torsional  simple harmonic damped oscillator driven by thermal fluctuations the response is\cite{Casimirbook}

\begin{equation}
S^2_{\rm ang} (f)= \frac{2k_{\rm B}T}{\pi\kappa Q f_{\rm r}}\frac{f_{\rm r}^4}{(f_{\rm r}^2-f^2)^2+f^2f_{\rm r}^2/Q^2}+ S^2_{\rm elec},
\label{power}
\end{equation}
\noindent where an independently determined  flat noise term $S^2_{\rm elec}$ associated with the electronic measurement setup\cite{decca4} has been added. $k_{\rm B}$ is Boltzmann's constant, $T$ the temperature at which the experiment is performed, $\kappa$ is the MTO's torsional constant. Doing the measurement at resonance, where the $1/f$ term and the detection noise are negligible, it is found that the minimum detectable force (per Hz$^{1/2}$) is 

\begin{equation}
F_{\rm min}=\frac{1}{b_i}\sqrt{\frac{2 \kappa_i k_{\rm B} T}{\pi Q_i f_{{\rm r}i}}} \sim 7 \frac{\rm fN}{\sqrt{\rm Hz}},
\label{Fmin}
\end{equation}

\noindent where the subindex $i$ stands for either Au or Ni. The drift in the resonant $f_{\rm r}$ is less than 5 mHz/hr under operating conditions.

\subsection{Separation and electrostatic calibration\label{sep}}

The general electrostatic calibration was performed similarly to what was done in Ref.~[\onlinecite{elcal}]. An optical fiber is rigidly attached to the MTO-sphere assemble, and a two-color interferometer was used to measure the distance between the assembly and the stationary engineered sample. Simultaneously $f_{\rm r}$ and the angular deviation of the MTO were recorded as the sphere is moved closer to the sample. From the change in $f_{\rm r}(z)$ the gradient of the interaction between the sphere and the plate can be obtained when a potential difference is applied between them. Comparing the separation dependence of the gradient of the interaction with that of the known sphere-plate interaction

\begin{eqnarray}
F_{\rm e} (z, V) & = & -2\pi \varepsilon_{0} (V - V_{\rm o})^2 \sum_{n=0}^\infty \frac{\coth(u) - n \coth(nu)}{\sinh(nu)} \nonumber \\
 & = & -2\pi \varepsilon_{0} (V - V_{\rm o})^2 \sum_{m=0}^7A_m q^{m-1},
\label{el2}
\end{eqnarray}

\noindent the unknown parameters of the system can be obtained. In Eq.~(\ref{el2}) $\varepsilon_{0}$  is the permittivity of free space (in SI units), $V$ is an applied potential to the sample (the sphere-oscillator assembly is always kept grounded) and $V_{\rm o}$ is a residual potential difference between the plate and the sphere, $u = 1+z/r_{\rm s}$, $A_m$ are fitting coefficients, and $q = z/r_{\rm s}$. While the full expression is exact, the series is slowly convergent, and it is easier to use the shown approximation developed in Ref.~[\onlinecite{fit}]. Using this approach torsional spring constants $\kappa_{\rm Au} = (1.15 \pm 0.01)\times 10^{-9}$~Nm/rad and $\kappa_{\rm Ni} = (9.98 \pm 0.06)\times 10^{-10}$~Nm/rad are obtained. In all cases investigated, $V_{\rm o}$ was of the order of a few mV, larger when the Ni sphere was used. For all samples and configurations used, $V_{\rm o}$ was checked to be position and time independent. As customary in these experiments, the differential measurements were performed with $V=V_{\rm o}$ to minimize beyond detection the electrostatic contribution. 

\begin{figure}[htbp]
	\vspace{-0.7cm}
	\centerline{\includegraphics[width=9cm]{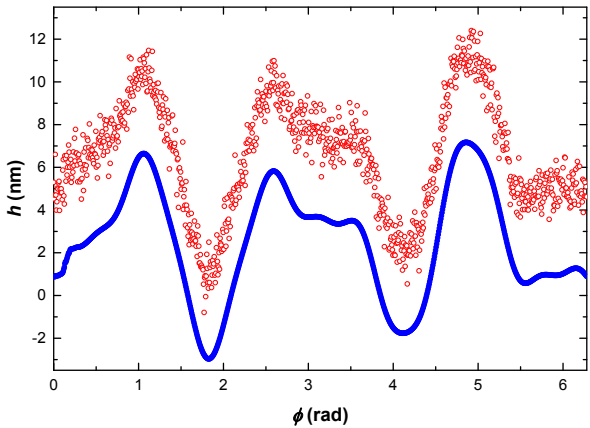}}
	\vspace{-0.3cm}
	\caption{(Color online) Measurement of the topography of the sample in a region where the measurements were done. Data extracted from WLI (\bl{$\bullet$}) and electrostatic measurements (\rd{$\circ$}), displaced up 4 nm for clarity.}
	\label{fig2a}
\end{figure}

In order to simplify the data acquisition and control of the system, during the experiment the two color interferometer is used such that it controls the separation between the sphere-MTO assembly and a fixed platform, instead of $z$.  Consequently, local variations of the height of the rotating sample are not taken into account in the calculation of $z$ while doing the experiment. It was verified, however, that these signals are attenuated below the experimental equipment sensitivity. The procedure used in the verification is the following: the topography of the sample $h(\phi)$ was extracted from the WLI, as shown in Fig. \ref{fig2a}. It was also verified that this topography corresponds, within $\delta z =$~0.6~nm, to the one obtained by using the experimental setup as an electric force microscope. Towards this end the resonant frequency of the sphere-MTO assembly was monitored as the sample was rotated at very low frequency at two different potential differences between the sphere and the rotating sample, at a separation $z = 200$~nm. The shift in the resonance frequency is associated with the gradient of the interacting force between sphere and sample, and the difference between the two measurements contains just the effect  of the gradient of the electrostatic force. From the electrostatic calibration this gradient can be converted into a separation, as shown in Fig. \ref{fig2a}.\cite{Casimirz} Similar results in the peak-to peak variation in $h(\phi)$ were obtained in all investigated samples. From $z(\phi)$ the change in the Casimir force is calculated (see Section \ref{calc} and Appendix A) and its component at the $f_r =n f$ is found to be negligible.

\section{Experimental data and errors\label{data}}

Data acquisition in the experiment requires a tight time and frequency synchronization. The time synchronization is given by a photolitographically defined sector located in $r\in[8.5,9.5]$~mm subtending an angle of $2\times10^{-4}$~rad. The leading edge of this sector is opposite the {\it cl}-line. In this region no Au is deposited. A diffraction limited laser is focused at $r\sim 9$~mm on the rotating sample and its reflection is measured by a photodiode. The edge on the change in reflectivity is detected and this defines the trigger for all timed events. It has been verified that this trigger  lags by $\tau_{\rm lag} = 10^{-6}/f$. The rotation frequency is obtained by monitoring $f_{\rm r}$ by finding the maximum of the thermally induced peak shown in Fig.~\ref{osc} with an accumulation time of 100 s. The required multiple of this signal is synthesized and fed to the air bearing spindle.

In general, with the sphere placed at $^{300}r_{\rm i} +75~\mu$m, the air bearing spindle was rotated at $\omega=2\pi f_{\rm r}/300$. In this manner, a force arising from the difference in the Casimir force between the metal coated sphere and the layered structure manifests itself at $f_{\rm r}$ even though  there are no parts moving at $f_r$. Using lock-in detection at $f_{\rm r}$ signals which are small but could show in conventional experiments are removed by the averaging provided by the rotating sample and the high-$Q$ of the MTO, as described in Section~\ref{exp}.  
\begin{figure}[htbp]
	
	\centerline{\includegraphics[width=8cm]{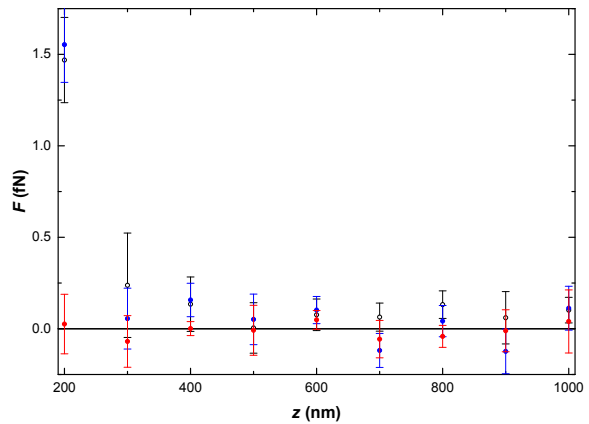}}
	\vspace{-0.3cm}
	\caption{(Color online) ($\circ$) Magnitude of $F(z)$ determined at $r=8.2$~mm. The signal is expected to be null in this situation. (\bl{$\bullet$}) signal measured in phase with the common {\it cl}-line, see Fig.~\ref{schem}. (\rd{$\bullet$}) Signal measured in quadrature.}
	\label{null}
\end{figure}

Before reporting on the data, better understanding of the system can be gained if the data is acquired when no signal is expected. In the case where the expected null result yields a measurable signal, then it is known that a systematic effect is present and needs to be subtracted. Fig.~\ref{null} show results obtained when the Au-coated sphere is placed at $r= 8.2$~mm, outside the outermost $n=300$ Au/Ni region. Unless otherwise stated, these data are representative and they represent an average of 3000 runs each with an integration time $\tau = 1$~s. It was verified that the measurements performed this way are consistent with a single measurement with $\tau = 3000$~s. All  repetitions for each measurement were confirmed to be consistent with a normal distribution and, consequently, the standard error of the mean was used as a good estimator for the experimental error at the 68\% confidence level. The results shown in Fig.~\ref{null} are similar to the ones observed in Ref.~[\onlinecite{DeccaPRL2014}]. 
The dependence of the observed signal with $\omega$ and $r$ has been investigated in Ref.~[\onlinecite{StPet}], and it was concluded it corresponds to an impulsive motion in $\theta(t)$ which happens once per revolution always along the same direction. For the purposes of this paper, it suffices to describe the methodology used to subtract this signal: the sample is rotated on the air bearing spindle until all the signal is observed in phase with the common {\it cl}-line, the red symbols in Fig.~\ref{null}. On the other hand, the blue symbols represent the  signal in quadrature and it is consistent with zero, as expected.  

\begin{figure}[htbp]
	\vspace{-0.7cm}
	\centerline{\includegraphics[width=9cm]{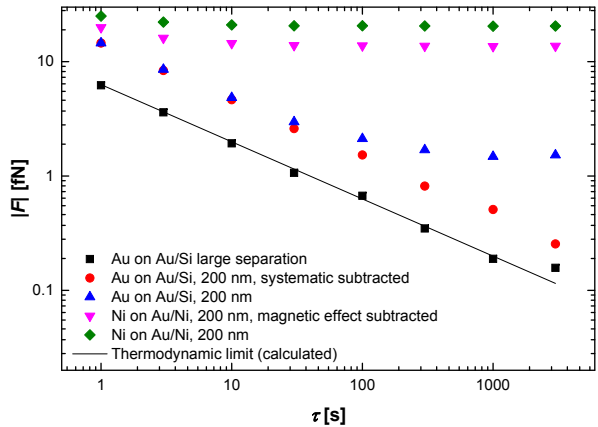}}
	\vspace{-0.3cm}
	\caption{(Color online) Time dependence of the magnitude of the lock-in detected signal for different samples and configurations.}
	\label{fig2b}
\end{figure}

Fig.~\ref{fig2b} shows the effect of time and spatial integration on the magnitude of the measured force. When the oscillator is far away from the rotating sample the signal (black squares) coincides, within the experimental error, with the calculated thermal noise (black line) from Eq.~(\ref{Fmin}).\cite{Casimirbook} For a sample with no expected Casimir signature where  Ni has been replaced by Si and $t$~=~200~nm, there is the remnant effect described in Fig.~\ref{null}. This can be observed at large $\tau$ on the blue triangles. When the systematic signal is subtracted as described above, the signal (red circles) shows a $1/\sqrt{\tau}$ decay as in random motion but with a larger magnitude than predicted by thermal noise alone. This increase of the random-like behavior is ascribed to a random fluctuation of the axis of rotation of the spindle.   These fluctuations in the axis of rotation produce a random, impulsive change in $z$ and manifests as an extra contribution to the effect of random thermal fluctuations. When a Ni coated sphere is used in front of the $t = 47$~nm sample a magnetic signature is observed. This signature (green diamonds) can be reduced by producing harmonic changes in the radial position of the sphere such that its harmonic frequency is non-commensurable with the frequency of rotation of the sample. Hence, after a sample full rotation a different magnetic configuration is underneath the sphere and a full spatial and temporal averaging is achieved (pink triangles). 

\begin{figure}[tbp]
	\vspace{-1.3cm}
	\centerline{\includegraphics[width=8cm]{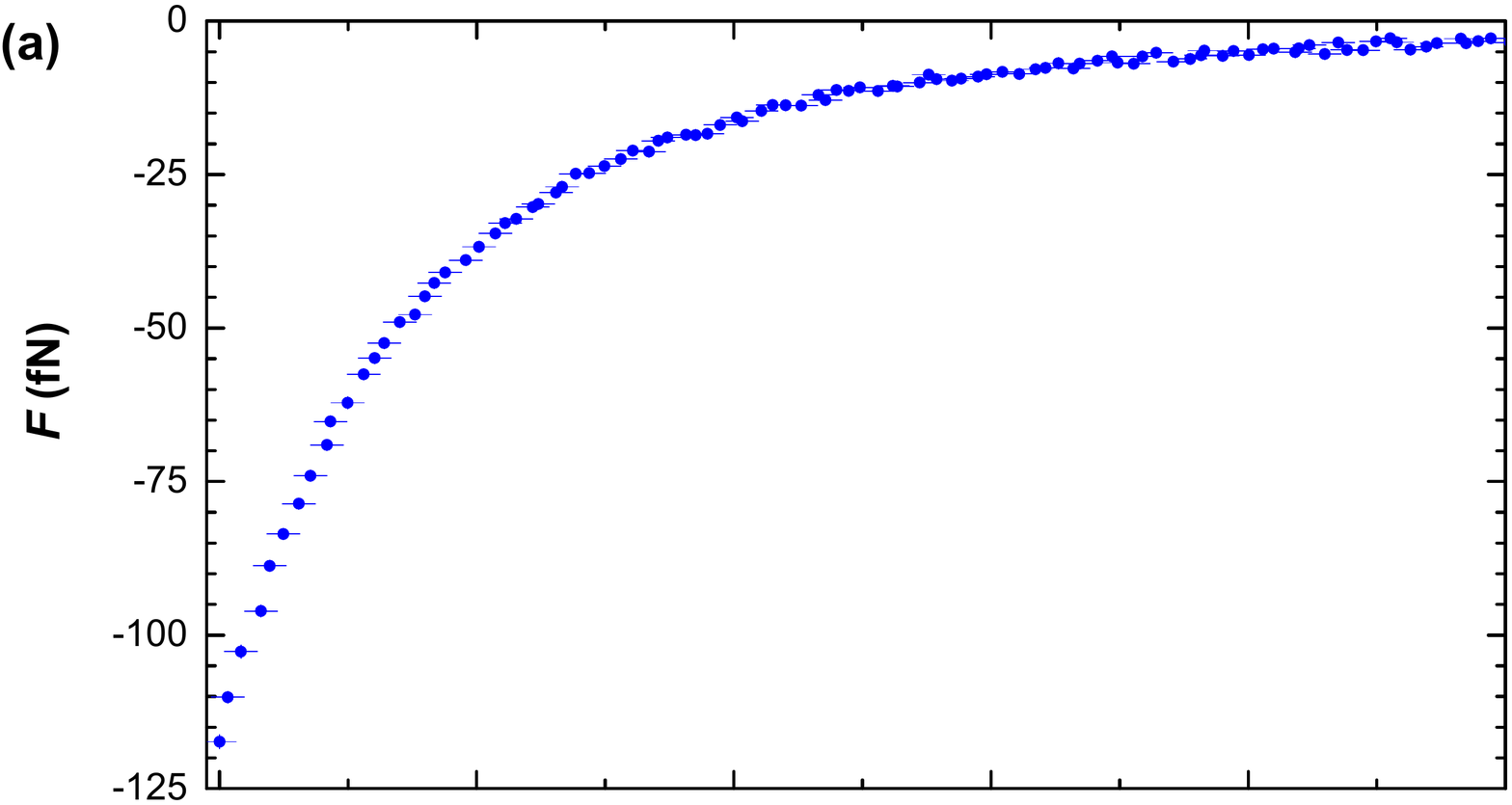}}
	\vspace{-2.5cm}
	\centerline{\includegraphics[width=8cm]{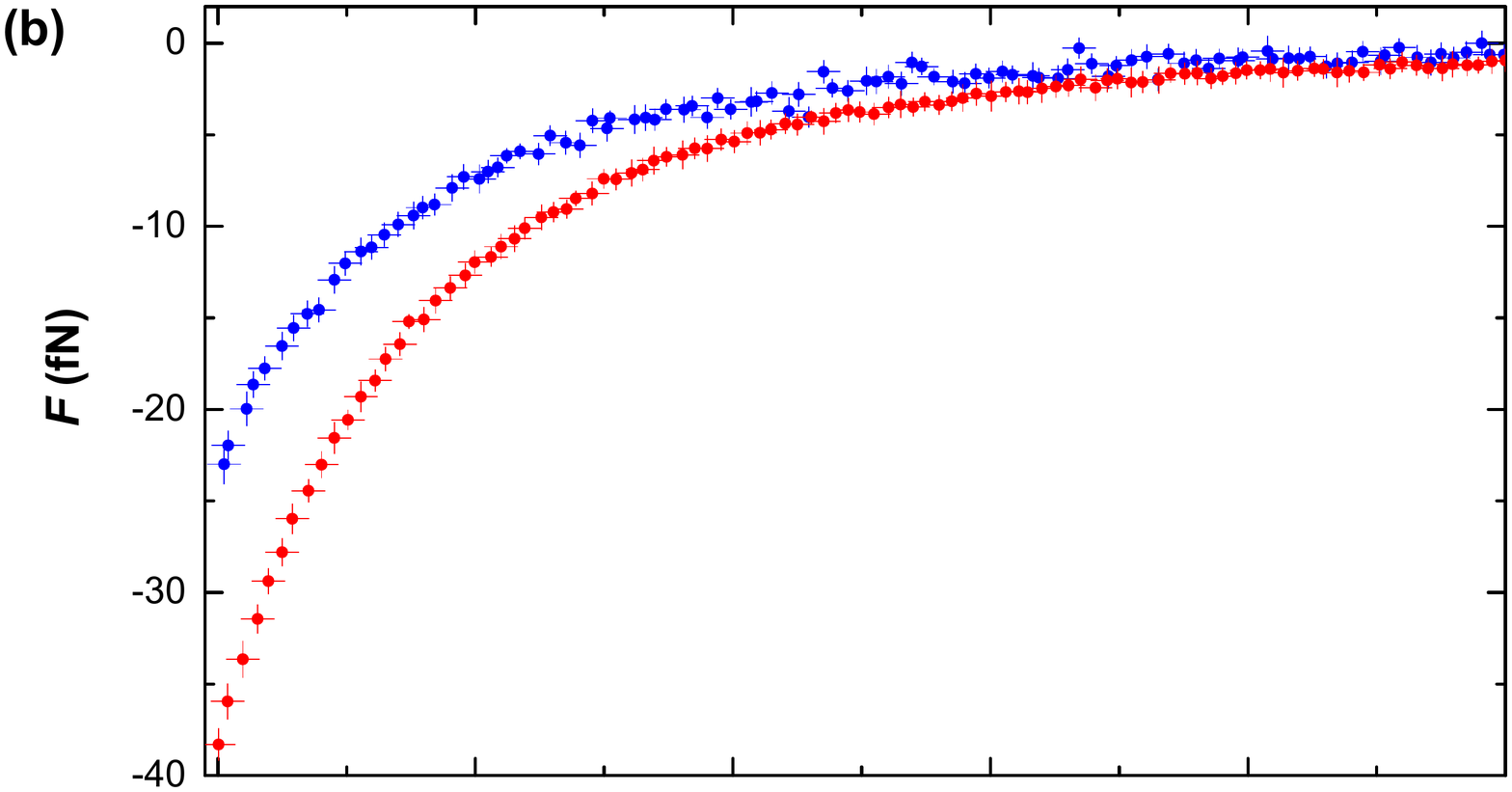}}
	\vspace{-2.5cm}
	\centerline{\includegraphics[width=8cm]{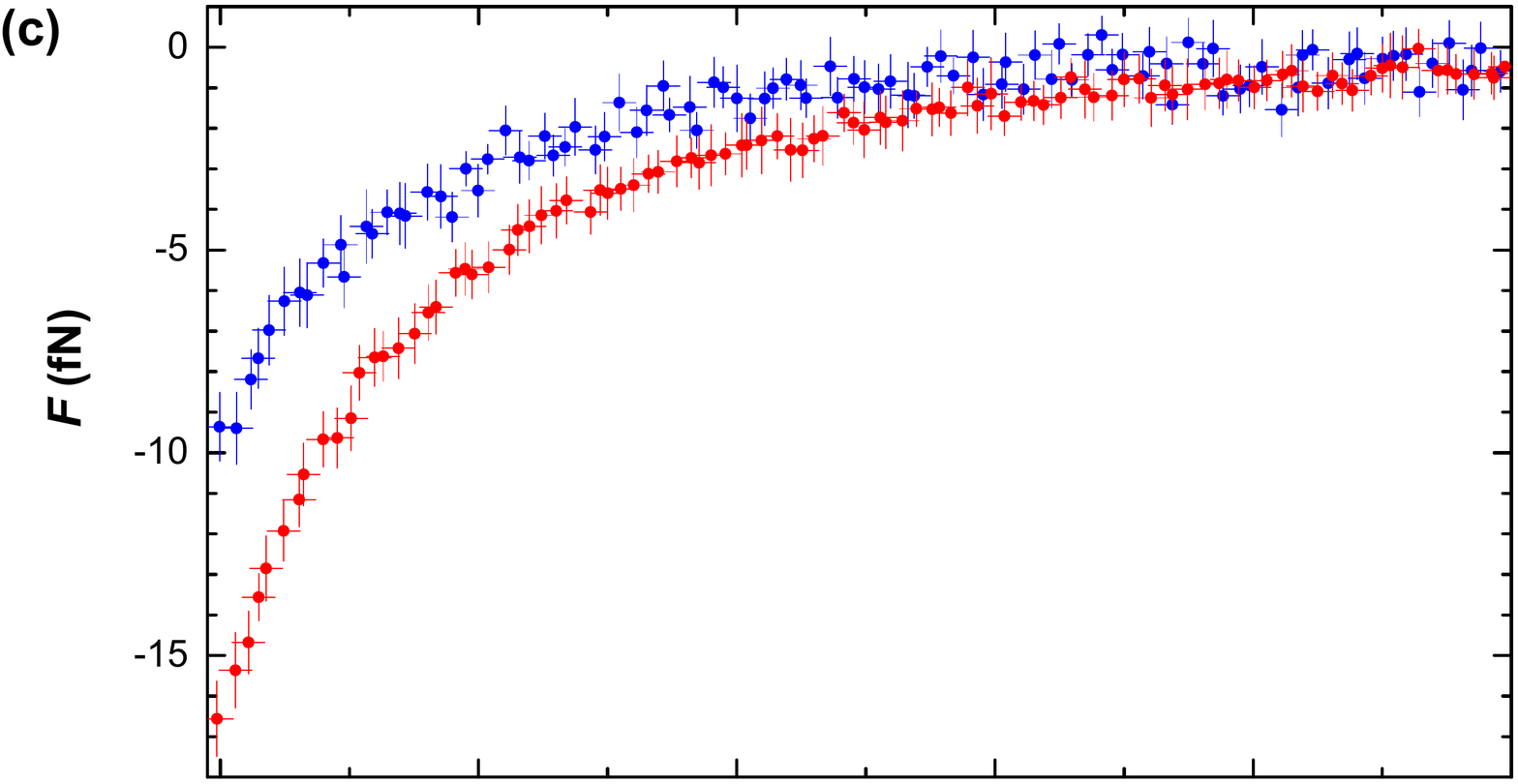}}
	\vspace{-2.5cm}
	\centerline{\includegraphics[width=8cm]{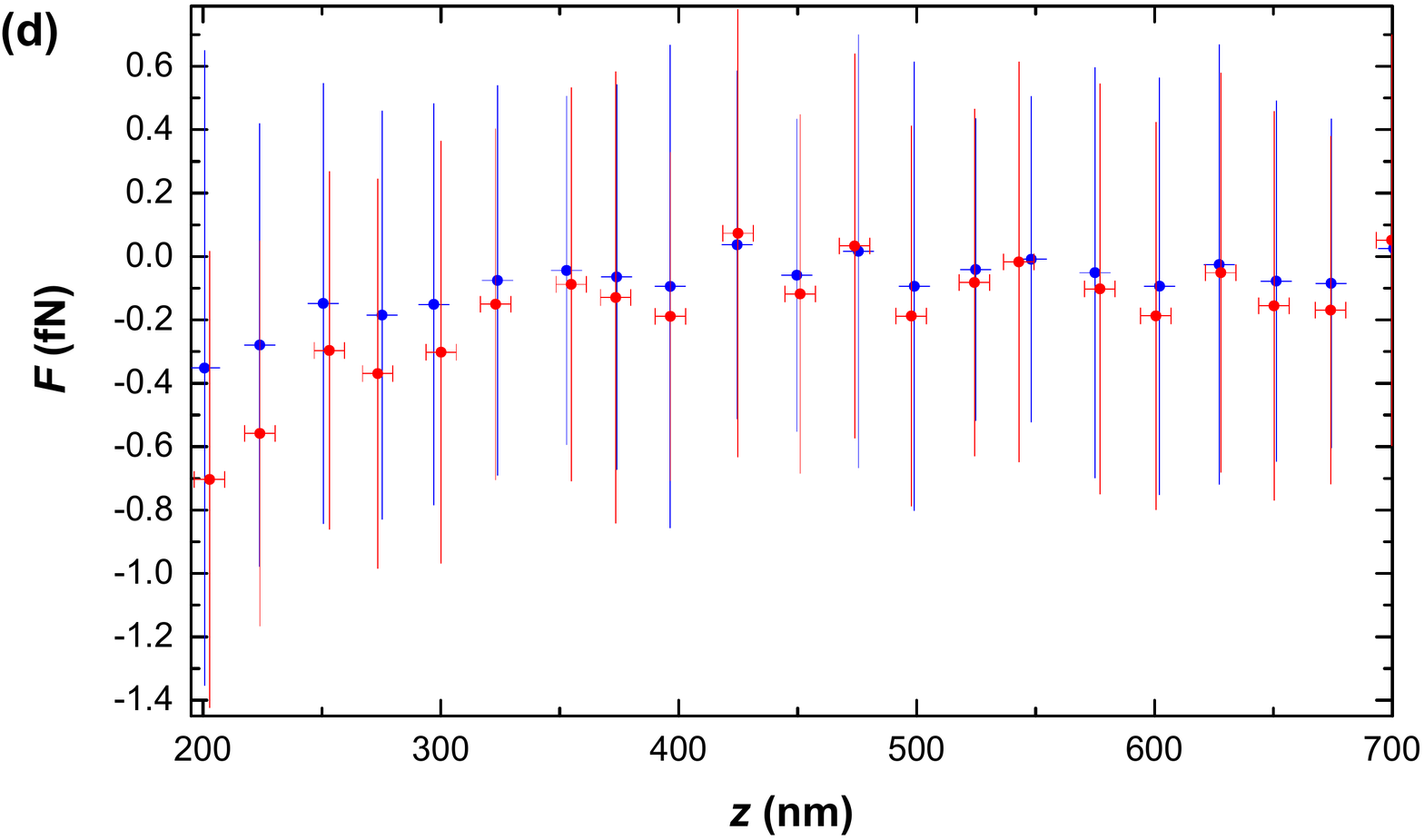}}
	\vspace{-0.5cm}
	\caption{(Color online) Measured force for  the Au- (red) and Ni-coated (blue) spheres as a function of the separation. (a) $t = 21$~nm sample. (Au-covered sphere results were not obtained) (b) $t = 37$~nm sample.  (c) $t = 47$~nm sample. (d) $t = 84$~nm sample. \\
	In all panels errors in the force include statistical and systematic errors at the 95\% confidence level. The error in the separation $\delta z = 2\times(\delta_{\rm m} z+ \delta_{\rm s} z)$  include the measurement ones $\delta_{\rm m} z = 0.6$~nm and the systematic error $\delta_{\rm s} z = 2.6$~nm obtained as the standard deviation from Fig.~\ref{fig2a}.}
	\label{datapl}
\end{figure}

Data collected in the experiment  from all samples ( $t =$~21, 37, 47, and 84~nm from top to bottom) are shown in Fig.~\ref{datapl}. Data for the interaction between the $t=21$~nm sample and the Au-coated sphere could not be collected due to a catastrophic failure of the system. Errors in the separation include measurement ones $\delta_{\rm m} z = 0.6$~nm and the systematic error $\delta_{\rm s} z = 2.6$~nm obtained as the standard deviation from Fig.~\ref{fig2a}.  Note that the results for the  $t=84$~nm sample can be used to place an upper bound on the magnitude of the magnetic remnant force. For a fixed separation $z+t+d_{\rm Ti}$ of the sphere tip from the Ni surface,  the magnetic remnant force is expected to be independent of the thicknesses $t$ and $d_{\rm Ti}$ of the gold and Ti caps.

\section{Calculation of the Casimir force\label{calc}}

In this Section calculations of the  Casimir force for the  experimental configuration are presented.   Polar coordinates  $(r,\phi)$ with origin  at the center of the sample are used on the sample surface. The origin of the angle $\phi$ is placed along a radius coinciding with one of the Ni-Au boundaries. Then, if  $n$ is the number of the periodically alternating Ni-Au regions (for the radius $r$ where measurements are done), the Au-Ni boundaries are placed at angles   $\phi=\phi_m$ with $\phi_m=m\,\pi/ n \;,m=0,1,\dots,2n-1$.  Let  $F_{\rm C}(z;r,\phi)$ the instantaneous Casimir force acting on  the sphere, as its tip is placed above   the point of the rotating sample of coordinates $(r,\phi)$,   $z$ being the sphere-sample separation.  Since the Au and Ti over-layers covering the Ni and Au regions of the sample are sufficiently thin (thinner than the skin depth) for the Casimir force  $F_{\rm C}(z;r,\phi)$ to ``feel" the difference between the underlying Ni and the Au regions,     $F_{\rm C}(z;r,\phi)$ depends non-trivially on $\phi$  and  is $\phi$-periodic with angular period $\delta=2 \pi/n$:  $F_{\rm C}(z;r,\phi)=F_{\rm C}(z;r,\phi+\delta)$.  When  the sample   rotates with angular frequency $\omega$,  the force on  the sphere becomes  time-dependent $F_{\rm C}(t)=F_{\rm C}(z;r,\omega t)$    (assuming that $z$ and $r$ do not change significantly in time). The angular periodicity of $F_{\rm C}(z;r,\phi)$ implies of course that $F_{\rm C}(t)$ varies periodically in time, with  frequency $f=\omega/2 \pi$. 
As aforementioned, see Eq.~(\ref{omega}), the sample's angular velocity $\omega$ is adjusted such that $f$ matches the resonance frequency $f_r$ of the MTO. 
Because of the high mechanical quality of the MTO,  the amplitude of the MTO  forced oscillations is proportional to the  Fourier coefficient  of frequency $f_r$ of the force $F_{\rm C}(t)$, all its higher harmonics being out of resonance. This motivates  the following definition of the measured force $F$:  
\be
F=- \frac{{\rm i} \,\omega}{2}  \int_0^{2 \pi /\omega} dt\; F_{\rm C}(z;r,  \omega\, t)\;e^{{\rm i} n\omega  t}\;.\label{meas0}
\ee
By replacing the time $t$ by the angle $\phi$ as integration variable, $F$ can be expressed as
\be
F= -\frac{{\rm i}}{2} \int_{0}^{ 2 \pi} d \phi \; F_{\rm C}(z;r, \phi)\;\ e^{{\rm i} n \phi}\;.\label{meas}
\ee
It is important to stress that the measured force $F$ represents a {\it differential} quantity,  probing the difference among the Casimir forces on the Ni and  Au sectors of the sample. The (angular)  average  $\langle {F}_{\rm C} \rangle$ of the sphere-sample Casimir force is automatically subtracted from $F$, and so are of course all other angle-independent   forces that may possibly act on the sphere. 
   
For the radius $r=^{300}r_{\rm i} + 75~\mu$m where most  measurements are performed, the lateral width $L=78.5\,\mu$m of the Ni-Au sectors is  much larger than the characteristic radius $\rho=\sqrt{R \,z} \simeq 5 \,\mu$m of the Casimir interaction region.  This implies that
for  most of the angles $\phi$ in Eq. (\ref{meas}), the (lateral) distance of the  sphere tip  from the closest of the Ni-Au boundaries is larger than $\rho$.  Because of this, the function $F_{\rm C}(z;r, \phi)$ is approximated by the step-function ${\hat F}_{\rm C}(z;r, \phi)$ defined such that:
\be
{\hat F}_{\rm C}(z;r, \phi)= F_{ \rm Au}(z) \chi(n \phi)+F_{ \rm Ni}(z) \chi(n \phi-\pi )\;.\label{step}
\ee  Here, $\chi(\theta)$ is the $2 \pi$-periodically continued  step-function of the interval $[0, 2 \pi[$, which is one for $0 \le \theta < \pi$, and zero elsewhere.   The forces  $F_{ \rm Ni}(z)$ and  $F_{\rm Au}(z)$   can be identified, respectively, with the Casimir forces  between the sphere and  two infinite homogeneous planar  slabs, one made of Ni and the other of Au, covered with  Au and  Ti over-layers.\cite{footnote} Thus, according to Eq. (\ref{step}) the force changes abruptly from $F_{ \rm Ni}(z)$ to  $F_{\rm Au}(z)$ (or viceversa), as the sphere tip crosses a Au-Ni boundary. 
Substituting Eq. (\ref{step}) into Eq. (\ref{meas}), 
\be
F(z)=F_{ \rm Au}(z)-F_{ \rm Ni}(z)\;.\label{meas2}
\ee
is obtained. The above formula makes fully explicit the differential character of $F$. Corrections to Eq. (\ref{meas2}) for edge effects due to the Au-Ni boundaries are considered at a later stage.

   

The forces $F_{ \rm Ni}(z)$ and  $F_{\rm Au}(z)$ are computed as follows.   Given the thickness $t_{\rm met}=250$ nm of the Au or Ni coatings of the sphere, it is  possible to model the sphere as a solid ball entirely made of either Ni or Au. Using the Proximity Force Approximation (PFA), the Casimir forces $F_{\rm Ni/Au}(z)$ can be expressed as:
\be
F_{\rm Ni/Au}^{(\rm PFA)}(z)=2 \pi R\,{\cal F}_{\rm Ni/Au}(z)\;,\label{PFA}
\ee
where $R$ is the sphere radius, and ${\cal F}_{\rm Ni}(z)$ ( ${\cal F}_{\rm Au}(z)$) denotes the free energy per unit area  of a homogeneous slab made of the same material as the sphere coating,   at distance $z$ from a planar   Au-Ti-Ni (Au-Ti-Au) three-layer slab, consisting of a  Au layer of   thickness $t$ followed by a layer of Ti of thickness  $d_{\rm Ti}$ covering an infinitely thick   Ni (Au) slab (given  the large thickness $d_{\rm tm}> 2\,\mu$m of the Au and Ni sectors,   the underlying Si substrate can be neglected).   The unit-area free energy ${\cal F}_{\rm Ni/Au}(z)$ can be estimated using the following generalization of  the famous Lifshitz formula to  layered slabs  consisting of an arbitrary number of magneto-dielectric layers:\cite{richmond,tomas}
$$
{\cal F}(T,a)=\frac{k_B T}{2 \pi}\sum_{l=0}^{\infty}\left(1-\frac{1}{2}\delta_{l0}\right)\int_0^{\infty} d k_{\perp} k_{\perp}  
$$
\be
\times \; \sum_{\alpha={\rm TE,TM}} \log \left[1- {e^{-2 a q_l}}{R^{(1)}_{\alpha}({\rm i} \xi_l,k_{\perp})\;R^{(2)}_{\alpha}({\rm i} \xi_l,k_{\perp})} \right]\;.\label{lifs}
\ee

In this equation $k_B$ is Boltzmann constant, $\xi_l=2 \pi l k_B T/\hbar$ are the (imaginary) Matsubara frequencies, $k_{\perp}$ is the modulus of the in-plane wave-vector, $q_l=\sqrt{\xi_l^2/c^2+k_{\perp}^2}$, and $R^{(j)}_{\alpha}({\rm i} \xi_l,k_{\perp})$ is the   reflection coefficient  of the possibly layered slab $j$ for polarization $\alpha$.    The extension of Lifshitz theory to magnetizable materials characterized by a dynamic  magnetic permeability $\mu(\omega)$ was developed by Richmond and Ninham \cite{richmond}.   Superscripts  $1$ and $2$ denote, respectively, the homogeneous slab and the three-layer system. Then  $R^{(1)}_{\alpha}$ coincides with the familiar Fresnel reflection coefficient of a homogeneous planar slab of  (dynamic) electric   permittivity $\epsilon_{1}$  and magnetic permeability $\mu_{1}$:
\be
R^{(1)}_{\rm TE}=\frac{\mu_{1}({\rm i} \xi_l) q_l-  \,k_l^{(1)}}{\mu_{1}({\rm i} \xi_l) q_l+ \,k_l^{(1)}}\;,\label{freTE}
\ee
\be
R^{(1)}_{\rm TM}=\frac{\epsilon_{1} ({\rm i} \xi_l) \,q_l- \,k_l^{(1)}}{\epsilon_{1}({\rm i} \xi_l) \,q_l+  \,k_l^{(1)}}\;,\label{freTM}
\ee
where $k_l^{(1)} \equiv k^{(1)}(\xi_l)$ and
\be
k^{(1)}(\xi) = \sqrt{\epsilon_{1}({\rm i} \xi) \mu_{1}({\rm i} \xi) \, \xi ^2/c^2+k_{\perp}^2}\;.\label{ka}
\ee  
For  the Ni-coated sphere $\epsilon_{1}({\rm i} \xi)=\epsilon_{\rm Ni}({\rm i} \xi)$, and $\mu_{1}({\rm i} \xi)=\mu_{\rm Ni}({\rm i} \xi)$, while for the Au-coated sphere $\epsilon_{1}({\rm i} \xi)=\epsilon_{\rm Au}({\rm i} \xi)$, and $\mu_{1}({\rm i} \xi)= 1$.
 
The expression for the reflection coefficient $R^{(2)}_{\alpha}$ of the  three-layer slab is more elaborate.\cite{born,Yeh} In the case of the Au-Ti-Ni slabs it reads:
\be  
R_{\alpha}^{(2)}({\rm i} \xi_l,k_{\perp})=\frac{r_{\alpha}^{(0{\rm Au})}+e^{-2\,t\, k_l^{({\rm Au})}}\,r_{\alpha}^{({\rm Au Ti Ni})}}{1+e^{-2\,t\, k_l^{({\rm Au})}}\,r_{\alpha}^{(0{\rm Au})}\,r_{\alpha}^{({\rm Au Ti Ni})}}\;,\label{three1}
\ee   
where
\be  
r_{\alpha}^{({\rm Au Ti Ni})} =\frac{r_{\alpha}^{({\rm Au Ti})}+e^{-2\,d_{\rm Ti}\, k_l^{({\rm Ti})}}\,r_{\alpha}^{({\rm Ti Ni})}}{1+e^{-2\,d_{\rm Ti}\, k_l^{({\rm Ti})}}\,r_{\alpha}^{({\rm Au Ti})}\,r_{\alpha}^{({\rm Ti Ni})}}\;,\label{three2}
\ee  
and 
\be
r^{(ab)}_{\rm TE}=\frac{\mu_{b}({\rm i} \xi_l) \,k_l^{(a)}-\mu_{a}({\rm i} \xi_l) \,k_l^{(b)}}{\mu_{b}({\rm i} \xi_l) \,k_l^{(a)}+\mu_{a}({\rm i} \xi_l) \,k_l^{(b)}}\;,\label{freTEgen}
\ee
\be
r^{(ab)}_{\rm TM}=\frac{\epsilon_{b}({\rm i} \xi_l) \,k_l^{(a)}-\epsilon_{a}({\rm i} \xi_l) \,k_l^{(b)}}{\epsilon_{b}({\rm i} \xi_l) \,k_l^{(a)}+\epsilon_{a}({\rm i} \xi_l) \,k_l^{(b)}}\;,\label{freTMgen}
\ee
where $k_l^{(a)} \equiv  k^{(a)}(\xi_l)$ with $k^{(a)}(\xi)$  defined as in Eq. (\ref{ka}),  $\epsilon_a$ and $\mu_a$ denote the electric and magnetic permittivities of medium $a$, and $\epsilon_0=\mu_0=1$ are used. The reflection coefficient for the Au-Ti-Au three-layer system is obtained by substituting  Ni by   Au everywhere in Eqs. (\ref{three1}-\ref{freTMgen}).

To apply Eqs. (\ref{lifs}-\ref{freTMgen}) for the calculation of the Casimir free energy, it is necessary to know the electric   permittivities $\epsilon_{a}({\rm i} \xi_l)$ of all materials (Au, Ti and Ni)  and the magnetic permittivity $\mu_{\rm Ni}({\rm i} \xi_l)$ of Ni, for $l$ large enough. One notes first   that, according to Eqs. (\ref{freTE}-\ref{freTMgen}), the reflection coefficients $R^{(1)}_{\rm TM}(0,k_{\perp})$ and  $R^{(2)}_{\rm TM}(0,k_{\perp})$ for TM polarization at zero frequency (i.e. for $l=0$)  are both equal to one in the setup, at it must be because  metallic surfaces screen out electrostatic fields.  Leaving aside for a moment the troublesome $l=0$ mode for TE polarization, consider the non-vanishing Matsubara modes with $l>0$.  For room temperature the frequency of the first Matsubara mode $\xi_1$ is about $10^{14}$ rad/s.  Since this   is very large compared to the relaxation frequency $\omega_{\rm rel}\sim 10^9$ rad/s  of the Ni magnetic permeability, it is clear that for all $l>0$    $\mu_{\rm Ni}({\rm i} \,\xi_l)=1$ can be used (the same is true of course for Ti and Au, which are non-magnetic). Thus, in order to estimate the $l>0$ terms in the sum Eq. (\ref{lifs}), one only needs to estimate the electric permittivities $\epsilon({\rm i}\, \xi_l)$ of the materials.   The standard procedure   is to compute $\epsilon({\rm i}\, \xi_l)$ using Kramers-Kronig relations, from tabulated values of ${\rm Im}\, \epsilon(\omega)$.\cite{palik} Since the latter are known only in a limited range of frequencies, especially on the low-frequency side,  it is  necessary to extrapolate the data for ${\rm Im}\, \epsilon(\omega)$ towards zero frequency, on the basis of some theoretical model.  This is usually done using the simple Drude model for ohmic conductors;
\be \epsilon_{\rm
Dr}(\omega)=1-\frac{\Omega^2}{\omega(\omega+{\rm  i}
\gamma)}\;,\label{drude}\ee where $\Omega$ is the plasma
frequency, and $\gamma$ is the relaxation frequency.
For the numerical computations  the  tabulated optical data of Au, Ni and Ti,\cite{palik} were used together with the  following values of the Drude parameters: $\Omega_{\rm Au}=8.9 \,{\rm eV}/\hbar$,  $\gamma_{\rm Au}=0.035\, {\rm eV}/\hbar$,\cite{book2}  $\Omega_{\rm Ni}=4.89 \,{\rm eV}/\hbar$,  $\gamma_{\rm Ni}=0.0436\, {\rm eV}/\hbar$, $\Omega_{\rm Ti}=2.42 \,{\rm eV}/\hbar$,  $\gamma_{\rm Ti}=0.034\, {\rm eV}/\hbar$.\cite{ordal} In order to rely as much as possible on the tabulated data and minimize the contribution of the Drude extrapolation, the values of $\epsilon({\rm i}\,\xi_l)$ were estimated by means of a recently proposed weighted form of Kramers-Kronig relation,\cite{generKK,generKK2} which strongly suppresses the contribution of low frequencies to the dispersion integral.

In what follows  the problematic contribution of the $l=0$ TE mode is taken into account. Determining the correct magnitude of this term for conductors has become an unresolved  puzzle. The problem is to  find the correct expression of the reflection coefficients $R^{(j)}_{\rm TE}(0,k_{\perp})$  to be inserted into the $l=0$ term of Eq. (\ref{lifs}). Surprisingly,
several precision experiments performed in recent years appear to rule out the physically natural prescription (dubbed as Drude prescription), according to which the reflection coefficients $R^{(j)}_{\rm TE}(0,k_{\perp})$ should be defined as the zero-frequency limit of the TE reflection coefficient of a ohmic conductor.   Instead, good agreement with these experiments is obtained if the  reflection coefficients $R^{(j)}_{\rm TE}(0,k_{\perp})$ are defined to be the zero-frequency limit of a {\it dissipationless} plasma model, with full neglect of relaxation processes of conduction electrons. 

\begin{figure}
\includegraphics [width=.9\columnwidth]{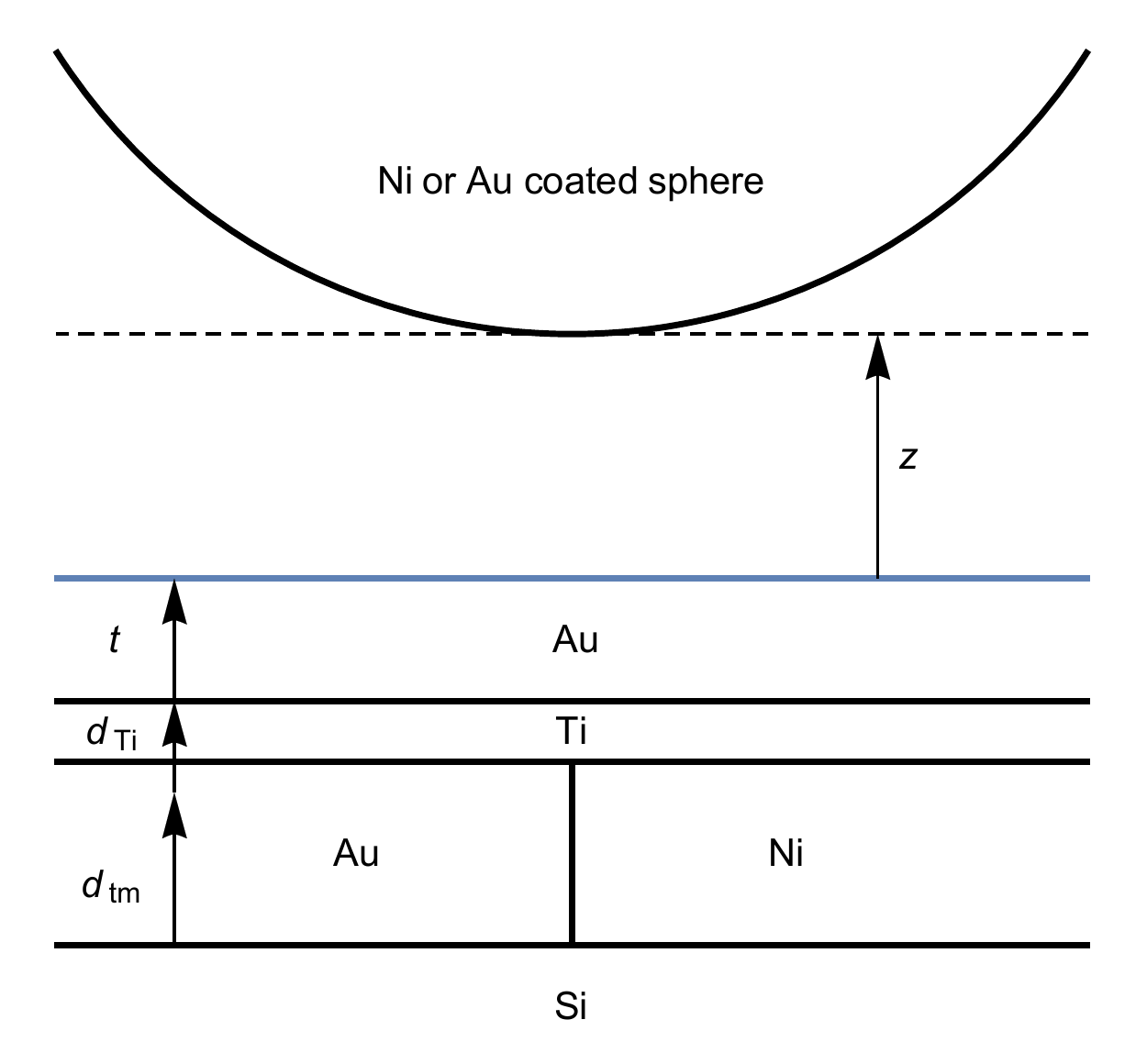}
\caption{\label{setup1}   Cross section (not in scale) of a small region of the rotating sample, showing two of the alternating Au-Ni sectors.  The Figure illustrates the layered structure of the sample.}

\end{figure}

Now consider the implications of the two prescriptions for the $l=0$ TE reflection coefficients in the experimental configuration. 
When the Drude prescription is used,  the $l=0$ TE reflection coefficient   $R^{(1)}_{\rm TE}(0,k_{\perp})$ of the homogeneous slab is found to be: 
\be
R^{(1)}_{\rm TE}(0,k_{\perp})|_{\rm Drude}=\frac{\mu_{1}(0)-1}{\mu_{1}(0)+1}\;.\label{refDr1}
\ee
On the other hand, for the  reflection coefficient  of the three-layer Au-Ti-Ni slab   one finds
\be
R^{(2)}_{\rm TE}(0,k_{\perp})|_{\rm Drude}=e^{-2 k_{\perp}\,(t+d_{\rm Ti})}\frac{\mu_{\rm Ni}(0)-1}{\mu_{\rm Ni}(0)+1}\;,\label{refDr2}
\ee 
while for the Au-Ti-Au slab
\be
R^{(2)}_{\rm TE}(0,k_{\perp})|_{\rm Drude}= 0\;.\label{refDr3}
\ee 
Since Lifshitz formula Eq. (\ref{lifs}) involves the product of the reflection coefficients of the slabs, it follows from Eqs. (\ref{refDr1}-\ref{refDr3}) that within the Drude prescription the $l=0$ TE mode contributes only to the Casimir force $F_{\rm Ni}(z)$ among the Ni-coated sphere and the Ni sectors of the rotating sample. Things are completely different with the plasma prescription. For the  reflection coefficient  $R^{(1)}_{\rm TE}(0,k_{\perp})$ of the homogeneous slab one gets:
\be
R^{(1)}_{\rm TE}(0,k_{\perp})|_{\rm plasma}=\frac{\mu_{1}(0) k_{\perp}-\sqrt{\mu_{1}(0) \, \Omega_1^2/ c^2 + k_{\perp}^2}}{\mu_{1}(0) k_{\perp}+\sqrt{\mu_{1}(0) \, \Omega_1^2/ c^2 + k_{\perp}^2}}\;.\label{refpl1}
\ee
This equation shows that within the plasma prescription, the $l=0$ TE reflection coefficient of  the homogeneous slab  is (in general) different from zero, for both the Au and Ni sphere coatings.
An analogous computation shows that  $R^{(2)}_{\rm TE}(0,k_{\perp})|_{\rm plasma}$is different from zero, both for the Au-Ti-Ni and the Au-Ti-Au three-layer slabs. The explicit expression of   $R^{(2)}_{\rm TE}(0,k_{\perp})|_{\rm plasma}$ will not be given here for brevity. The conclusion is that within the plasma prescription, the $l=0$ TE mode does contribute  both to $F_{\rm Au}$ and $F_{\rm Ni}$, for both  sphere coatings.  The different  values of the $l=0$ TE reflection coefficients engendered by the Drude and the plasma prescriptions   imply a huge difference between the respective predictions of the force $F$  measured using the Ni-coated sphere. Before   this is shown, it is  opportune to  examine various corrections that need to be considered.

Equation  (\ref{PFA}) was obtained using the PFA, and thus it is not exact. Recently, curvature corrections to PFA have been worked out  by several authors.\cite{fosco2,grad1,grad2,teo} According to these works, the exact Casimir force $F_{\rm C}$ between a sphere of large radius and a planar surface can be expressed as
\be
F_{\rm C}(z)=F_{\rm PFA}(z)\;\left[1+\theta(z) \frac{z}{R}+ o\left(\frac{d}{R}\right) \right]\;,\label{corrPFA}
\ee
where $F_{\rm PFA}(z)$ is the PFA result given in Eq. (\ref{PFA}). For the sphere-plate system, the coefficient $\theta$ has been estimated by the authors of Ref~[\onlinecite{fosco2,grad1,grad2,teo}] for a variety of cases, including  both  perfect conductors and  real metals, for zero temperature as well as for finite temperatures. In all cases it has been found that for submicron separations $z$,  the coefficient $\theta$ is negative and that its absolute value is less than one. Importantly, it has been found that $\theta$ is only weakly dependent on  detailed material properties of the conductor considered,  like its plasma frequency, relaxation frequency or temperature.   Because of that it is possible to estimate curvature corrections for any metallic plates at room temperature using the value of $\theta$  for perfect conductors at $T=300$ K, that was computed in Ref.~[\onlinecite{grad2}]. 
Eq. (\ref{corrPFA}) can be used   to  correct for curvature effects the  differential force $F$ in Eq. (\ref{meas2}).  
Curvature corrections vary from less than 0.1 $\%$ (for $z=200$ nm) to less than 0.3 $\%$ (for $z=500$ nm).

The correction due to (small scale) surface roughness is examined next. The root-mean-squared surface roughnesses of the sphere $\delta_{\rm S}$, and sample $\delta_{\rm P}$     were determined by means of  AFM scanning, and were found to be $\delta_{\rm S}= 1$nm and $\delta_{\rm P}=$ 3 nm. Since $\delta_{\rm S}$ and $\delta_{\rm P}$ are both small compared to the separations, the  roughness correction can be estimated by the multiplicative approach.\cite{book2}  By using this procedure 
\be
F_{\rm R}=F \left[1+6 \frac{\delta_{\rm S}^2+\delta_{\rm P}^2}{z^2}+15 \frac{\delta_{\rm S}^4+6\,\delta_{\rm S}^2 \, \delta_{\rm P}^2+ \delta_{\rm P}^4}{z^4} \right]\;
\ee   
is obtained  for the roughness corrected force. The above equation implies that the roughness correction to the measured force $F$ varies from  0.15 $\%$ for $z=200$ nm, to 0.03 $\%$ for $z=500$ nm.   Since curvature corrections to PFA and roughness corrections have opposite signs, and thus tend to cancel each other, the combined  effect of curvature and roughness is smaller than their individual effects.


In what follows, edge effects arising from the Au-Ni boundaries are computed. In writing Eq. (\ref{step}) it was assumed that the Casimir force changes abruptly from $F_{\rm Au}(z)$ to $F_{\rm Ni}(z)$ when the (projection onto the sample surface of the) sphere tip moves from a Au  sector to a Ni sector of the sample, or viceversa. Of course, this is only an approximation. In reality one expects a smooth transition taking place in a narrow strip $\Sigma$ having a  width comparable to the interaction radius  $\rho=\sqrt{R\,z}$  on either side of the boundary.  In Appendix B  an estimate of the Casimir force in the transition region $\Sigma$ based on the Derjaguin approximation is derived. It is found that the stepwise approximation of Eq. (\ref{step}) is very good  for lateral displacements $y$ of the sphere tip from the closest Au-Ni boundary  such that $|y|> \rho$, while for $|y|<  \rho$ $F_{\rm C}$ is well approximated by a linear function of $y$ interpolating between   $F_{\rm Au}(z)$ and $F_{\rm Ni}(z)$. Using this more accurate expression of the Casimir force in the transition region 
\be
F =[F_{\rm Au}(z)-F_{\rm Ni}(z)] \left(1-\frac{\pi^2 z R}{6 L^2} \right)\; 
\ee 
is obtained for the edge-effects corrected force. Consequently, the correction due to edge effects varies from 0.8 \% for $z=200$ nm  to 2 \% for $z=500$ nm.

Consider now the effect of  the topography of the sample surface.  Eq. (\ref{meas0}) implicitly assumed that the separation $z$ is constant. In reality, the surface of the rotating sample is not exactly planar (see Fig.~\ref{fig2a}). Its topography can be described by  a height profile $h(r, \phi)$, which varies slowly over the scale $\rho$ of the Casimir interaction area. The reference plane with respect to which $h$ is measured is fixed such that   $h(r, \phi)$ has zero angular mean $\langle h \rangle=0$   for the value of $r$ where  measurements are taken. 
In Appendix A it is shown that for a  small amplitude ($|h(r,\phi)| \ll z$) height profile,   the   force correction $\delta F$ is
\be
\delta F(z) =  -\langle F_{\rm C} \rangle   \, \frac{3 \, {\rm i}\, h_n}{2 z}  + 6\, F(z)  \,  \frac{ \langle h^2 \rangle  }{z^2}    \;,\label{heicor}
\ee
where
\be
  \langle F_{\rm C} \rangle = (F_{\rm Au}(z) +F_{\rm Ni}(z))/2
\ee 
is the average Casimir force, $h_n=\int_0^{2 \pi} d \phi \,h(\phi) \exp({\rm i} n \phi)$ is the $n$-th Fourier coefficient of $h(\phi)$, and $\langle h^2 \rangle$ is the angular average of $h^2(\phi)$. Using this formula, it is estimated that   the correction $\delta F$ has the small magnitude
\be
|\delta F(z) | < 10^{-6}  |\langle F_{\rm C} (z) \rangle| +10^{-4} |F(z)|\;,
\ee
and therefore it can be  neglected.

In our computations we neglected corrections from spatial dispersion. Based on the analysis carried out in \cite{raoul}, it can be expected that for the thicknesses of our metallic layers and for the sphere-plate separations (larger than 200 nm) that we consider, the correction to the force $ F$ due to spatial dispersion is smaller than 0.2 \%, and therefore it is negligible.

So far, it has been assumed that the instantaneous Casimir force $F_{\rm C}(z;r,\phi)$ for a rotating sample is not influenced by the relative speed between the sphere and the sample.  Strictly speaking this is not quite right, because the Casimir force between two surfaces in relative sliding motion depends on their relative velocity.\cite{volokitin} However, it turns out     
that the velocity-dependence of the Casimir force is totally negligible for  speeds $v$ such that $v/(2 z) \ll \omega_{\rm rel}$, where $\omega_{\rm rel}$ denotes the smallest frequency scale characterizing the electromagnetic response of the plates. In the setup, the slowest  time scale is set by the spin relaxation time in the Ni regions of the plates,  which has a characteristic frequency   $\omega_{\rm rel} \sim 10^9$ rad/s.\cite{lucy}   With
$\omega \simeq 2 \pi $ rad/s, $r < 1$ cm and $z> 200$ nm,    $v/( 2 z)=    \omega\, r/(2 z) < 2 \times 10^5$ rad/s is obtained. Since the slow motion condition  $v/(2 z) \ll \omega_{\rm rel}$ is well satisfied, neglecting  velocity effects is justified.

\section{Comparison between experiment and theory.\label{comp}}

In this section the theoretical errors in the force $F$ are estimated. The main sources of theoretical errors in the experiment are the uncertainties in the optical data of  Ni, Au and Ti, in the thicknesses $t$ and $d_{\rm Ti}$ of the Au and Ti layers, and in the sphere-sample separation $z$.  When estimating the theoretical error a conservative 10 \% uncertainty in   the electric permittivities $\epsilon({\rm i} \xi_n)$ of the three metals was allowed for  all non-vanishing Matsubara frequencies. A 10 \% uncertainty was also assumed for  the squares of the respective plasma frequencies $\Omega$. An uncertainty $\delta t= 1$ nm and $\delta t_{\rm Ti}= 1.5$ nm in the thicknesses of the Au and Ti over-layers, respectively, and an uncertainty $\delta z=1$ nm in the separation $z$ were considered.  The total theoretical error $\Delta F$ was computed at 68 \% confidence level, by combining in quadrature the individual theoretical errors.

\begin{figure}[htbp]
\includegraphics [width=.9\columnwidth]{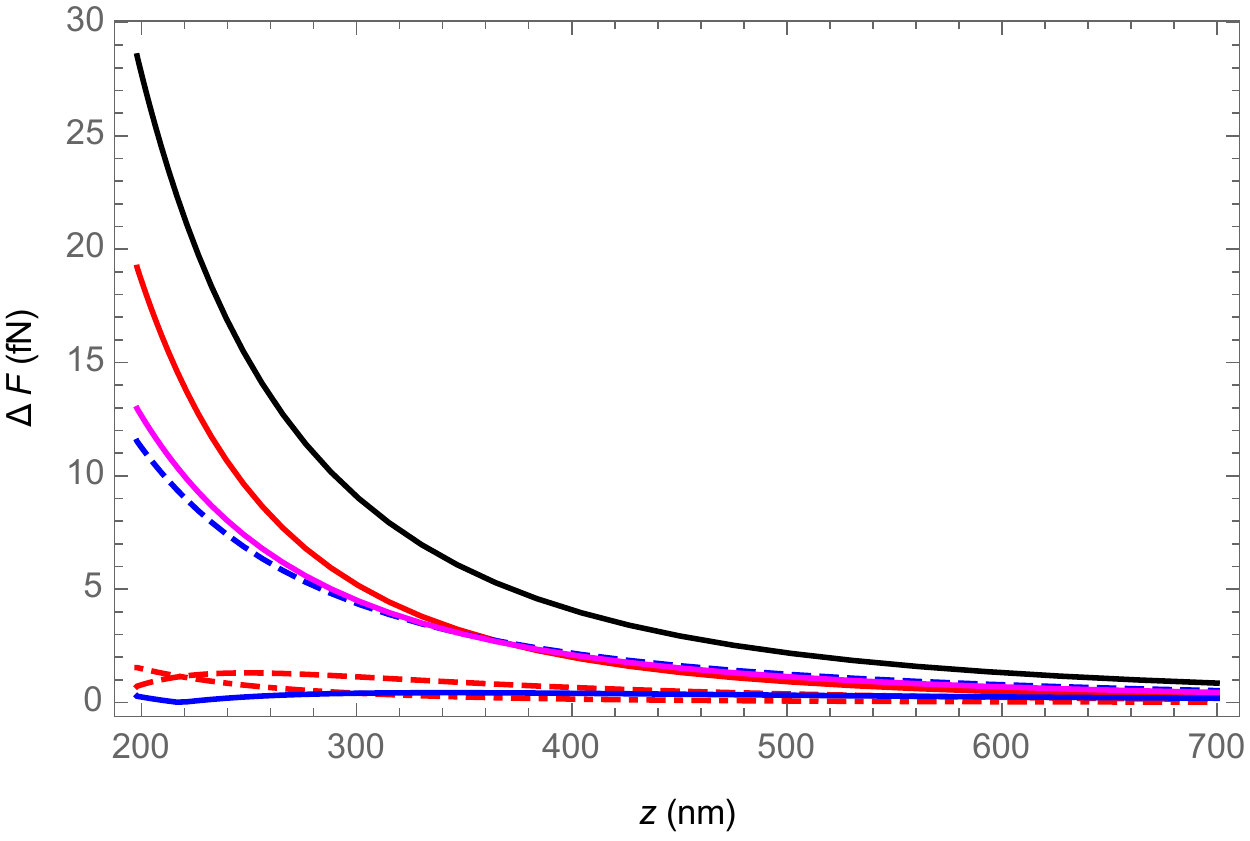}
\caption{\label{errNiDrnm37}  Theoretical errors in the force $F$ for the Ni-coated sphere opposed the 21 nm sample,  due to inaccuracy of optical data of Ni (solid red line), Au (red dashed line), Ti (red dot-dashed line), to the uncertainty $\delta t$ in the thicknesses of the Au layer (solid blue line),  to the uncertainty $\delta t_{\rm Ti}$ in the thicknesses of the Ti layer (dashed blue line), and to errors $\delta z$ in the separation (magenta curve). The total theoretical error is shown by the solid black line. The errors were computed using the Drude model, with inclusion of Ni magnetic properties. }
\end{figure}

\begin{figure}[htbp]
\includegraphics [width=.9\columnwidth]{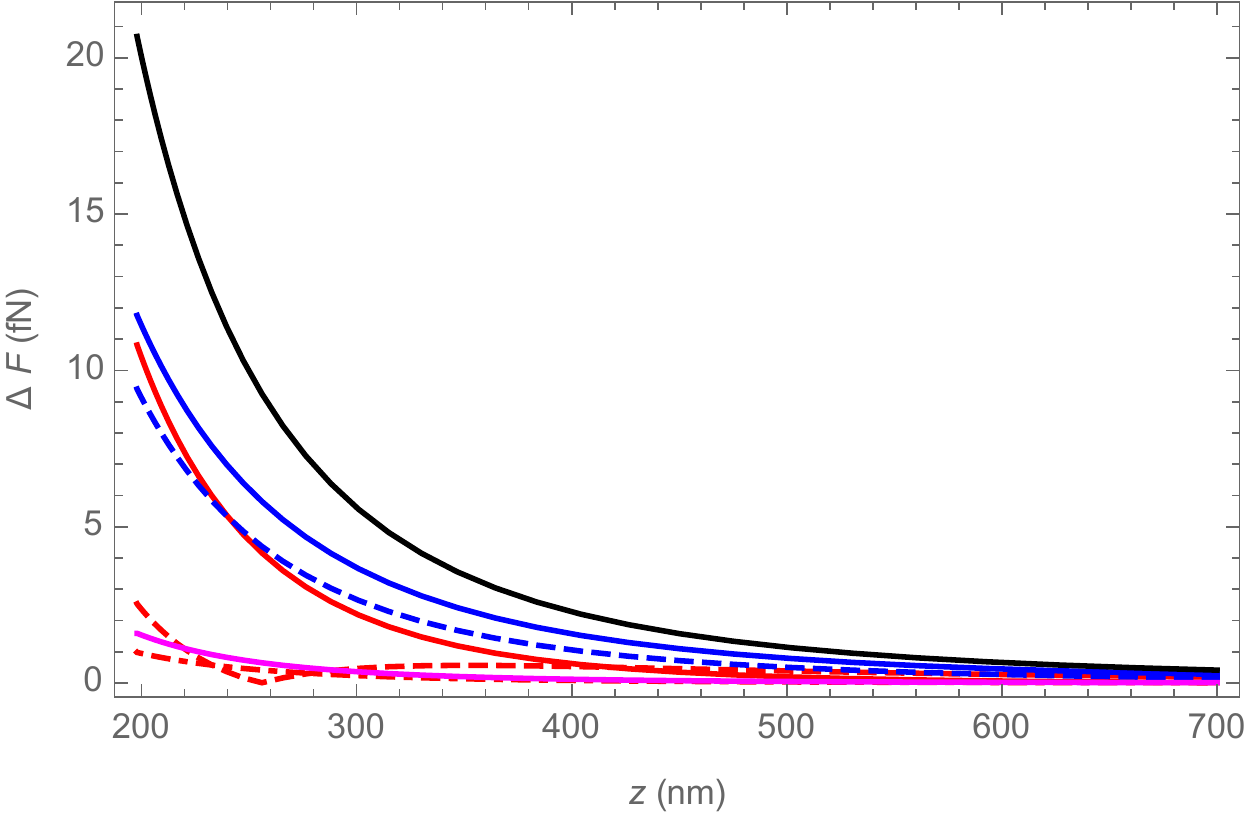}
\caption{\label{errNipl37}  Theoretical errors in the force $F$ for the Ni-coated sphere opposed the 21 nm sample,  due to inaccuracy of optical data of Ni (solid red line), Au (red dashed line), Ti (red dot-dashed line), to the uncertainty $\delta t$ in the thicknesses of the Au layer (solid blue line),  to the uncertainty $\delta t_{\rm Ti}$ in the thicknesses of the Ti layer (dashed blue line), and to errors $\delta z$ in the separation (magenta curve). The total theoretical error is shown by the solid black line. The errors were computed using the plasma model, with inclusion of Ni magnetic properties. }
\end{figure}

As an example,  Figs. \ref{errNiDrnm37} and  \ref{errNipl37}  show plots of the theoretical errors $\Delta F$ (in fN)  for the Ni-coated sphere opposed the 21~nm sample. Fig. \ref{errNiDrnm37}  was computed using  the Drude model, while Fig. \ref{errNipl37} was computed using the plasma model. In  both cases magnetic properties of Ni were included. The errors obtained for the Au sphere, or for either sphere but without taking into account magnetic properties of Ni  have magnitudes comparable to those displayed in Fig. \ref{errNipl37}.

\begin{figure}[htbp]
	\includegraphics [width=.9\columnwidth]{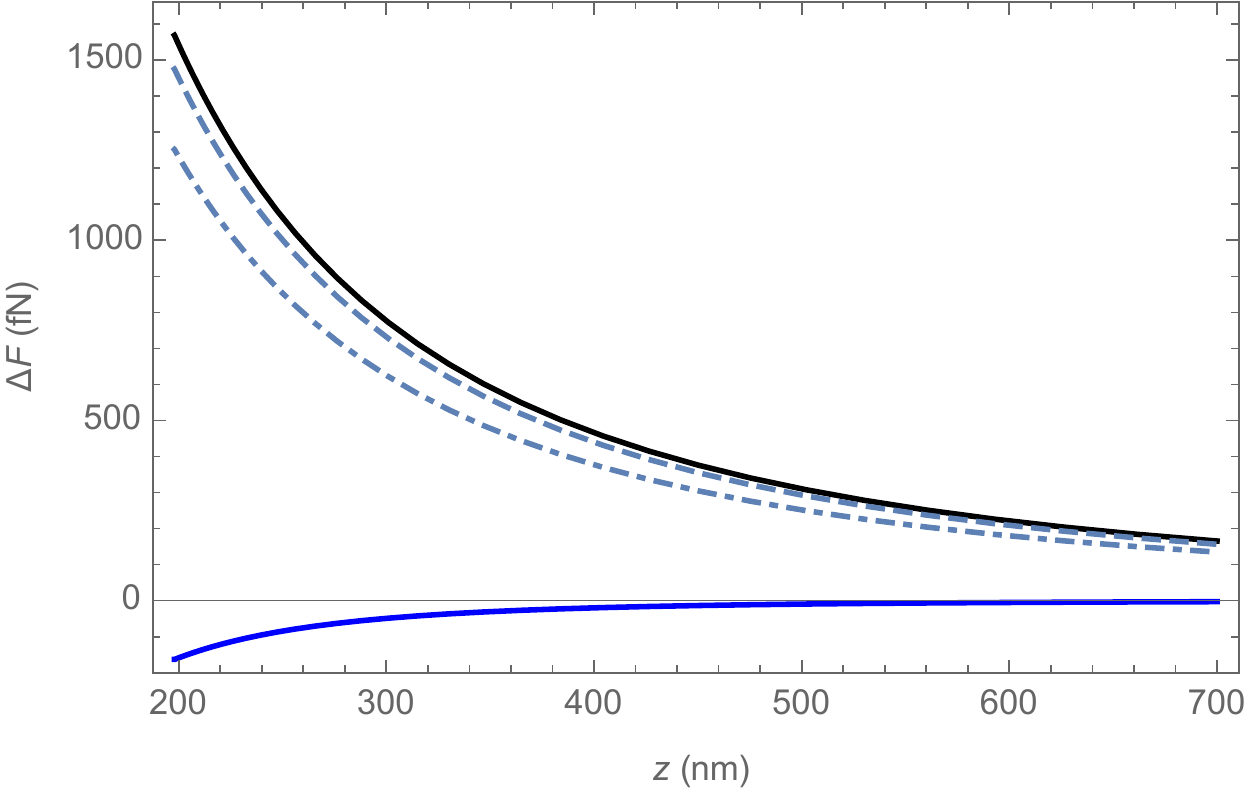}
	\caption{\label{muNi}  Force $F$ for a Ni sphere opposed the 21 nm sample predicted by  the Drude model  for different values of the static magnetic permeability  of Ni. The solid black line, the dashed and dot-dashed lines were computed using $\mu_{\rm Ni}(0)=110, 50$ and 20, respectively. The solid blue line corresponds to taking $\mu_{\rm Ni}(0)=1$. The latter model is referred to as the non-magnetic Drude model.}
\end{figure}

\begin{figure}
\includegraphics [width=.9\columnwidth]{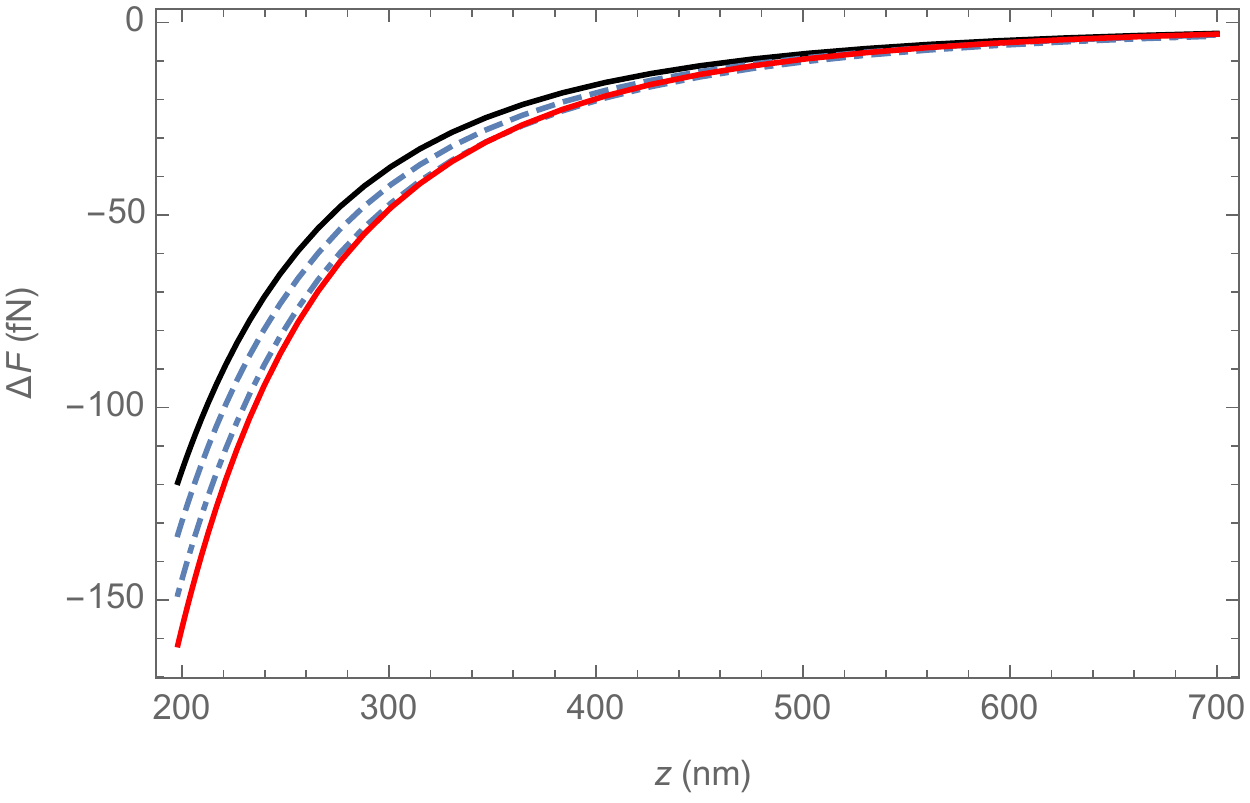}
\caption{\label{muNipl}  Force $F$ for a Ni sphere opposed the 21 nm sample predicted by  the plasma model  for different values of the static magnetic permeability  of Ni. The solid, dashed and dot-dashed lines were computed using $\mu_{\rm Ni}(0)=110, 50$ and 20, respectively. The red solid line corresponds to $\mu_{\rm Ni}=1$. The latter model is referred  to as the non-magnetic plasma model.}
\end{figure}

The large  force predicted by the Drude model is weakly dependent on the value of the static magnetic permeability of Ni, provided that $\mu_{\rm Ni}(0)$ is significantly larger than one. The force $F$ predicted by the Drude model for the Ni sphere opposed the 21 nm sample (using for the static magnetic permeability of Ni the three values $\mu_{\rm Ni}(0)=110$ (solid line), $\mu_{\rm Ni}(0)=50$ (dashed line) and $\mu_{\rm Ni}(0)=20$ (dot-dashed line)) are plotted in Fig. \ref{muNi}. For comparison, Fig. \ref{muNipl} shows the force predicted by the plasma model, again for a Ni sphere opposed the 21 nm sample, using for the static magnetic permeability of Ni the same three values $\mu_{\rm Ni}(0)=110$ (solid line), $\mu_{\rm Ni}(0)=50$ (dashed line) and $\mu_{\rm Ni}(0)=20$ (dot-dashed line). Also shown (red solid line) is the force predicted if magnetic properties of Ni are completely neglected, i.e. taking $\mu_{\rm Ni}(0)=1$.

\begin{figure}[htbp]
	\vspace{-0.8cm}
	\centerline{\includegraphics[width=9cm]{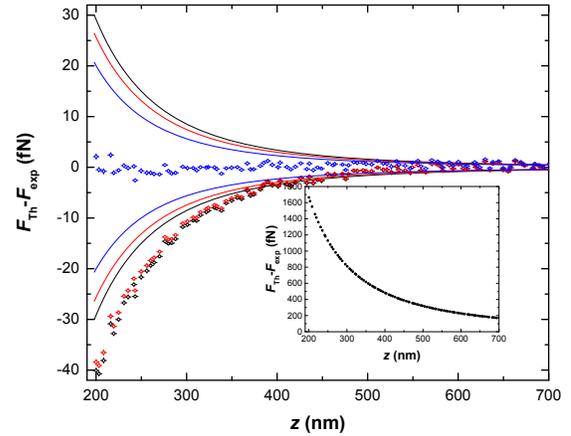}}
	\vspace{-0.5cm}
	\caption{(Color online) Difference between the theoretical and experimental determination for the forces as a function of separation for the $t=21$~nm sample. Three different models are used for the theoretical calculations: non-magnetic Drude (black),  non-magnetic plasma (red) and magnetic plasma (blue). In all cases the interaction is measured with the Ni-coated sphere.\cite{break} The inset shows the difference between the magnetic Drude model and the experimental data. All errors are at the 68\% level.}
	\label{21}
\end{figure}

\begin{figure}[htbp]
	\vspace{-0.8cm}
	\centerline{\includegraphics[width=9cm]{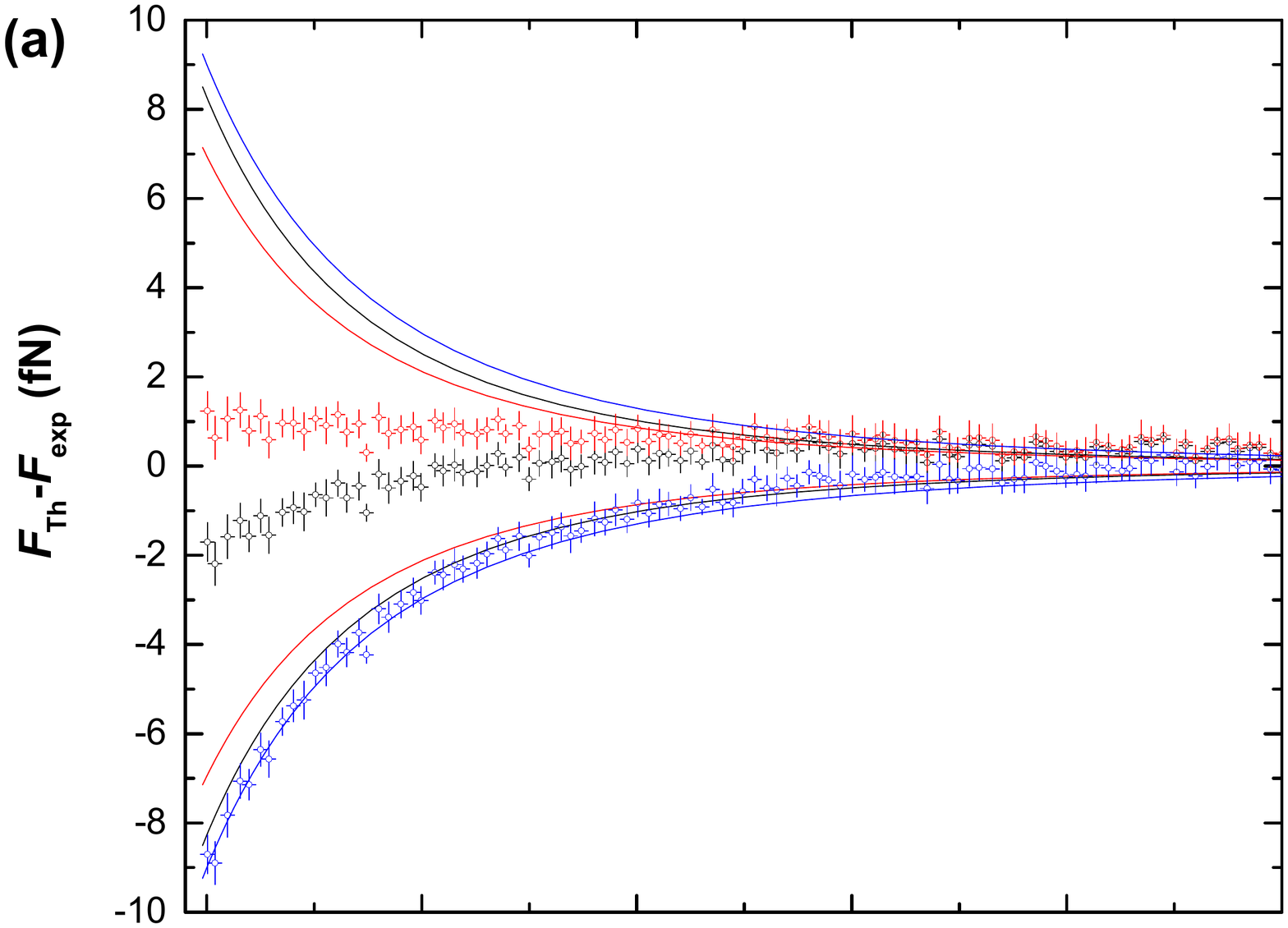}}
	\vspace{-2cm}
	\centerline{\includegraphics[width=9cm]{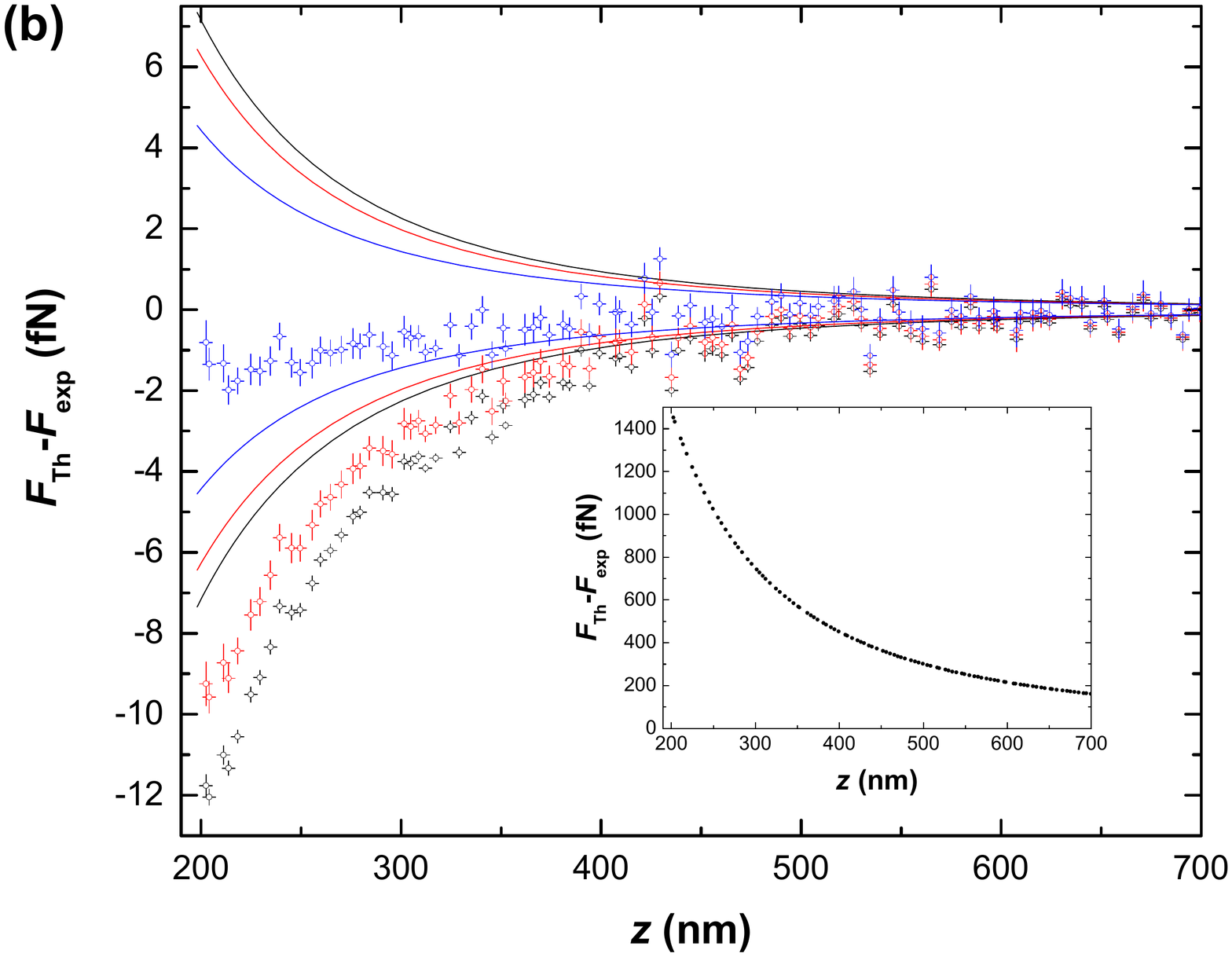}}
	\vspace{-0.5cm}
	\caption{(Color online) Difference between the theoretical and experimental determination for the forces as a function of separation for the $t=37$~nm sample. Three different models are used for the theoretical calculations: non-magnetic Drude (black),  non-magnetic plasma (red) and magnetic plasma (blue). (a) Situation when the Au-coated sphere is used. (b) Situation when the Ni-coated sphere is used. The inset shows the difference between the magnetic Drude model and the experimental data. All errors are at the 68\% level.}
	\label{37}
\end{figure}

\begin{figure}[htbp]
	\vspace{-0.8cm}
	\centerline{\includegraphics[width=9cm]{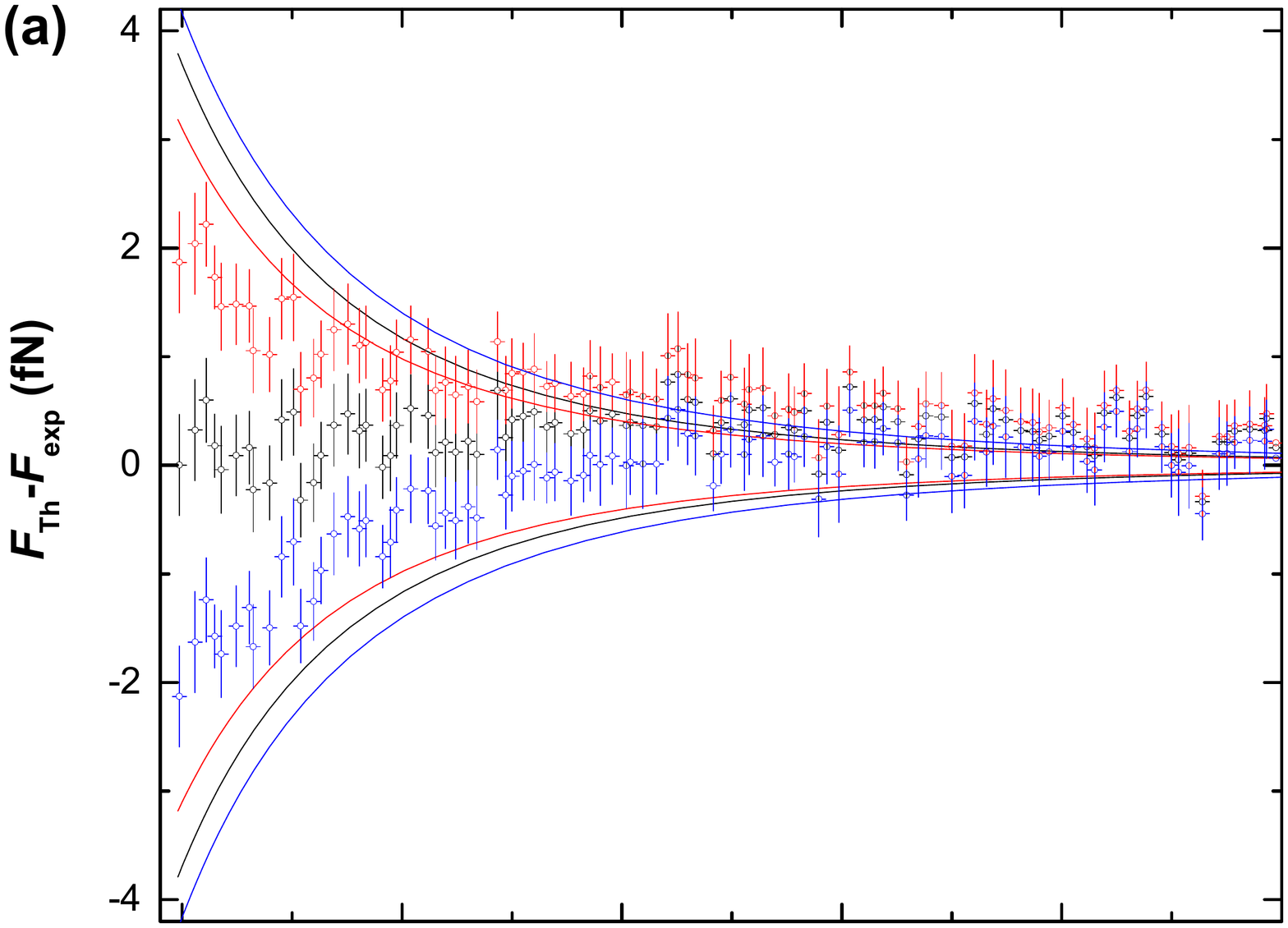}}
	\vspace{-2cm}
	\centerline{\includegraphics[width=9cm]{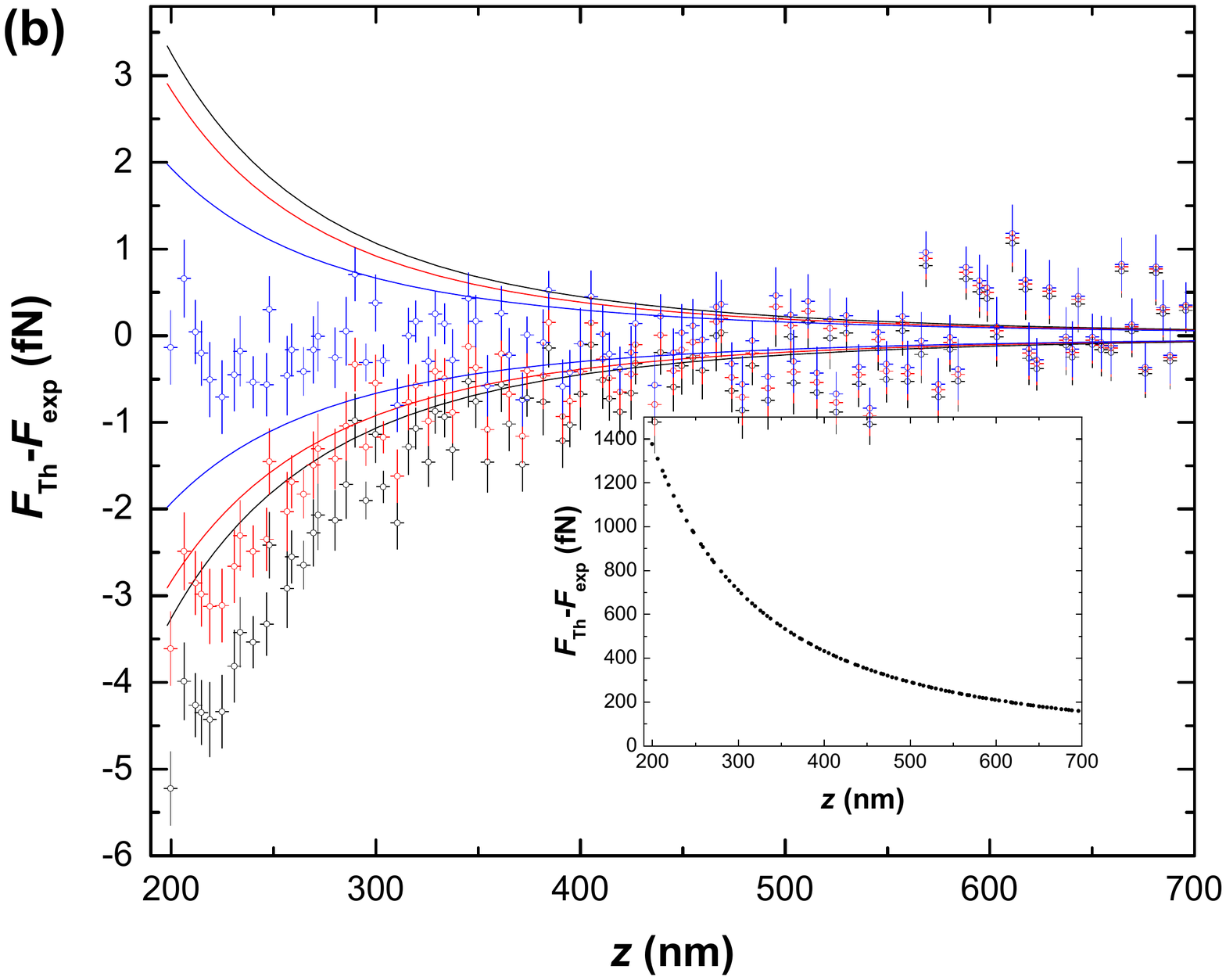}}
	\vspace{-0.5cm}
	\caption{(Color online) Difference between the theoretical and experimental determination for the forces as a function of separation for the $t=47$~nm sample. Three different models are used for the theoretical calculations: non-magnetic Drude (black),  non-magnetic plasma (red) and magnetic plasma (blue). (a) Situation when the Au-coated sphere is used. (b) Situation when the Ni-coated sphere is used. The inset shows the difference between the magnetic Drude model and the experimental data. All errors are at the 68\% level.}
	\label{47}
\end{figure}

In order to  assess if the data reveal or not the influence of the magnetic properties of Ni,  the data shall be compared below with four different theoretical models, i.e. the Drude and plasma models with and without inclusion of the Ni magnetic properties. The Drude and plasma magnetic models shall use $\mu_{\rm Ni}(0)=110$, while the non-magnetic models shall use $\mu_{\rm Ni}(0)=1$. Figs.~\ref{21} to \ref{47} show the difference between the calculated force $F_{\rm th}$ and the measured one $F_{\rm exp}$ for the samples with $t=$21, 37, and 41~nm. For the last two cases the force obtained when using a Ni- or Au- coated sphere are shown. In all cases the theoretical calculations are done using the magnetic plasma model ($\mu_{\rm Ni}=110$), and the non-magnetic versions of plasma and Drude model ($\mu_{\rm Ni}=1$). For comparison it is also shown that when the magnetic properties and dissipation are taken into account (magnetic Drude model) the differences between the calculations and the experimental data are over two orders of magnitude larger than the calculated errors. The case for $t=87$~nm is excluded because, except for the magnetic Drude model, all other models' predictions agree with the data. 

For the remaining three models (i.e. non-magnetic Drude (lossy) model, and the magnetic and non-magnetic versions of the plasma lossless model) the analysis of the data provides many revealing facts. 

We will first discuss the comparison between data and experiment when the magnetic plasma model is used. In this case the model agrees with the data for all investigated situations when 68\% confidence level  errors are used, and consequently cannot be ruled out by the experiment. In particular, the agreement is excellent for all the samples considered when the Ni-coated sphere is used. In these cases the difference between theory and experiments are nearly  indistinguishable from zero. The situation is different when the Au-coated sphere is used. For the $t = 37$~nm sample the agreement is marginal when the Au-coated sphere is used, but within the error bars. For the $t = 47$~nm sample the agreement is very good when considering the experimental errors in the measurements and in the theoretical calculations.  

The remaining two models (non-magnetic plasma and non-magnetic Drude models), saving small quantitative differences, produce quite similar results. They both provide a very good description of the data when the Au-coated sphere is used,  but they are excluded by the data at the 68\% confidence level for $z < 450$~nm when the interaction is measured using the Ni-coated sphere.

\begin{figure}[tbp]
	\vspace{-1.cm}
	\centerline{\includegraphics[width=9cm]{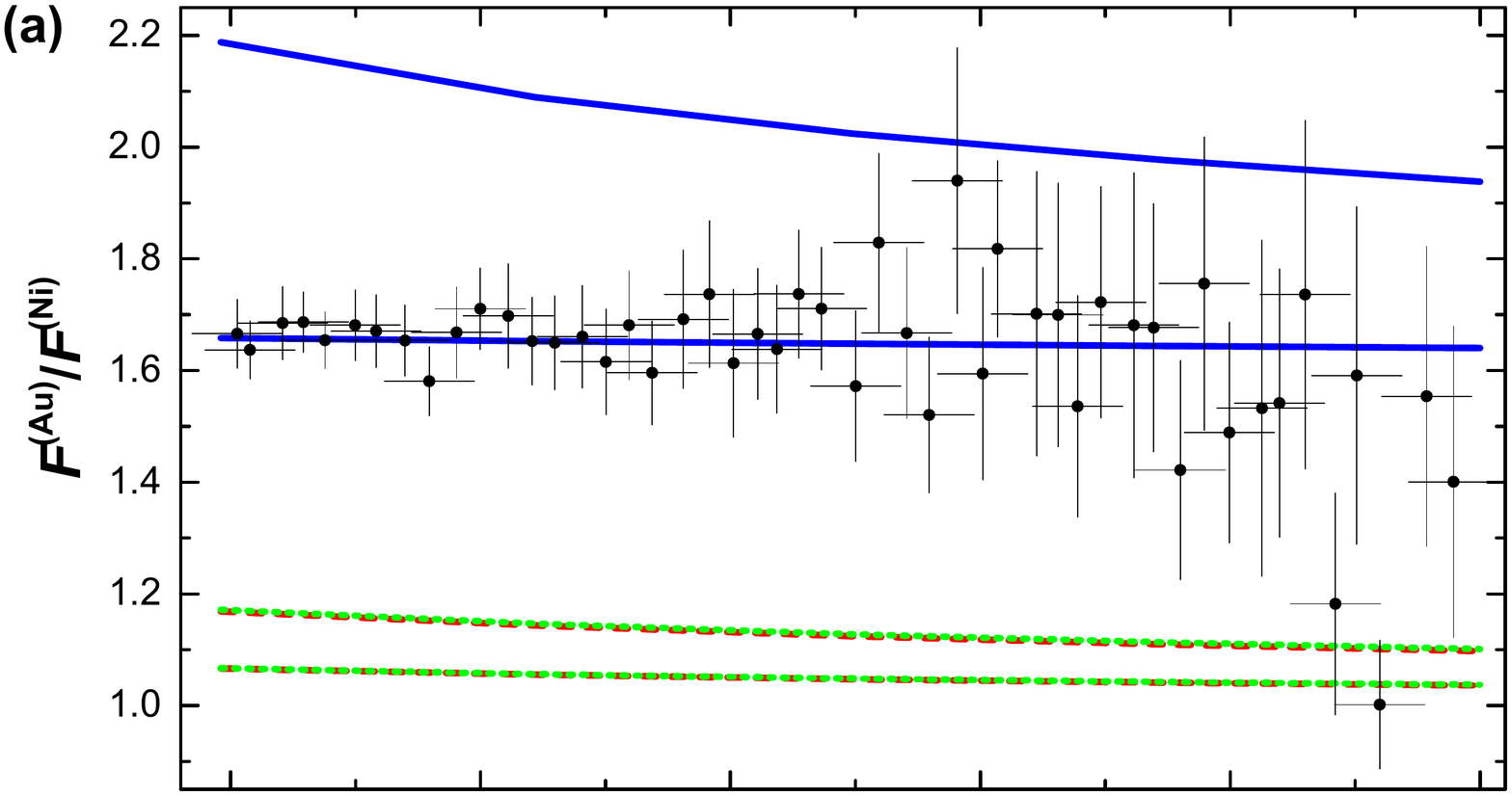}}
	\vspace{-2.8cm}
	\centerline{\includegraphics[width=9cm]{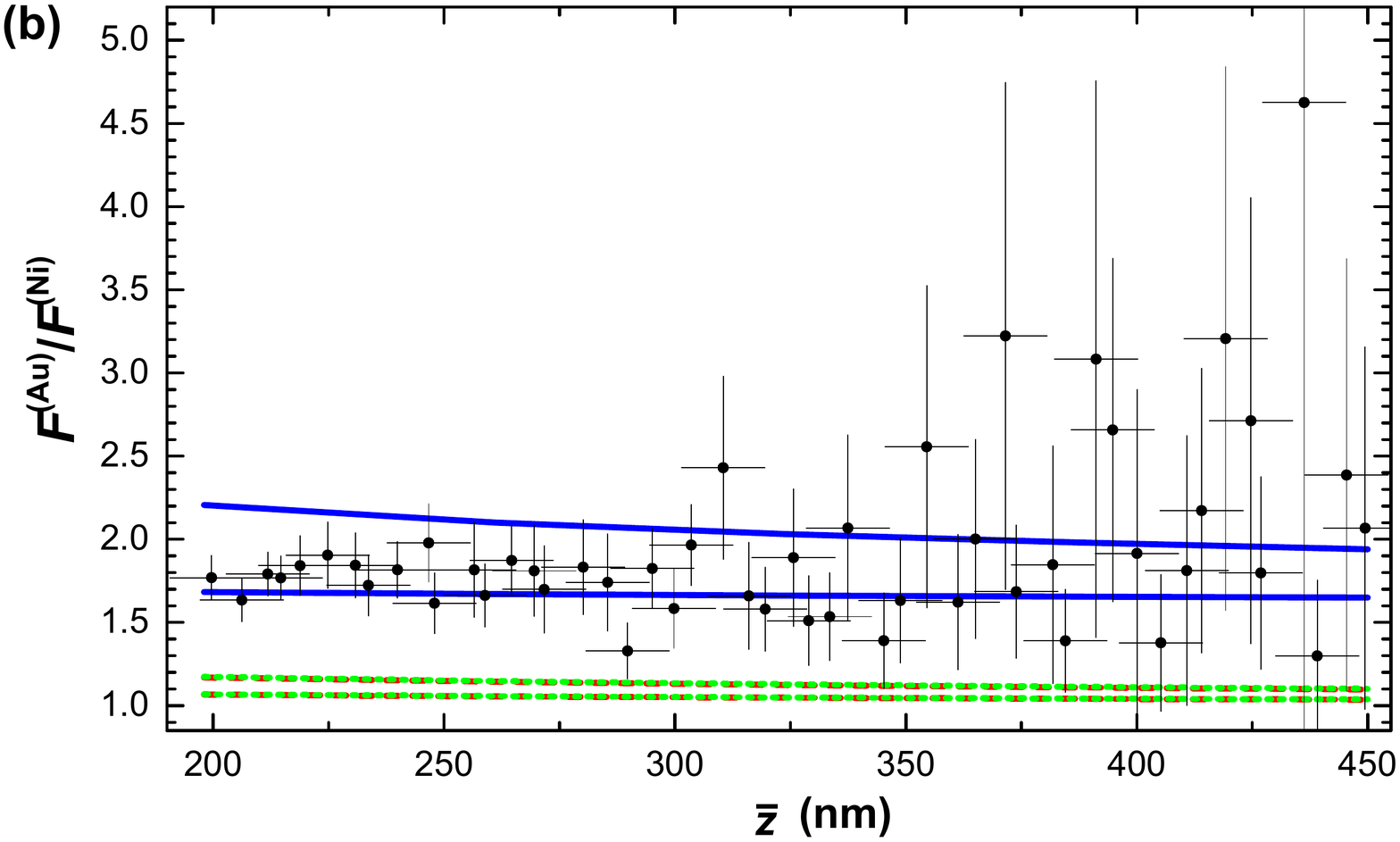}}
	\vspace{-1cm}
	\caption{(Color online) Ratio between the measured signal with the  Au-coated  and Ni-coated   spheres as a function of average separation $\bar{z}$. Data above $\bar{z} = 450$~nm have a large error. The lines enclose the 95\% confidence interval when variations of the experimental parameters are considered. Plasma with magnetic properties (solid, blue), plasma non-magnetic (dotted, green) and Drude non-magnetic (dashed, red). Predictions from the last two models are almost coincident. (a) $t = 37$~nm sample. (b) $t = 47$~nm sample.}
	\label{fig4}
\end{figure}

With the intent of further elucidating which of the remaining models provide the best description of the data, experiments using the $t = 37$~nm and $t = 47$~nm samples are analyzed while trying to minimize the effect of experimental uncertainties. Bearing in mind that all errors associated with the rotating samples themselves are the same independently of the sphere used, a more direct comparison with the models can be done if the ratio $F^{(\rm Au)}/F^{(\rm Ni)}$ between the signals measured with the Au- and Ni-coated spheres is considered. Fig.~\ref{fig4} shows the experimental ratio (plotted at the average separation $\bar{z} = (z^{(\rm Au)} + z^{(\rm Ni)})/2$ between the two experimental runs) and the calculated intervals at 95\% confidence for the ratio when the three different models are used. The calculated 95\% confidence bands were found by using the same experimental uncertainties as before.

\section{Conclusions\label{concl}}

In this paper a series of differential measurements in the Casimir regime have been performed. The increased sensitivity of the differential technique is based in the  superior suppression of all forces not associated with the spatial compositional variation of the sample, in particular patch potentials.  Experimental approaches to minimize magnetic contributions, as well as the effect of systematic impulsive motion of the motor used to produce the rotations were implemented. A judicious selection  of the overlaying metallic layer thickness allows to practically isolate the contribution of the $l=0$ TE model in the multilayer structure.  

An extensive analysis of errors and their effect in the comparison between the experimental results and theoretical calculations was presented. The largest sources of errors to perform the experimental-theory comparison arose from the uncertainties in the physical parameters on the sample. The effect of these uncertainties were partially taken into account by measuring the interaction between the same spatially inhomogeneous rotating sample and two spheres coated with Au and Ni. 

For all samples and separations investigated, it is unequivocally concluded that a magnetic Drude model cannot be used as a viable representation of a metallic sample in the presence of vacuum fluctuations. 

Of the remaining three models the situation is different for different samples. For the $t= 37$~nm sample the non-magnetic plasma  and non-magnetic Drude models are excluded by the data at the 95\% confidence level for separations $z<400$~nm. Similarly, for the $t=47$~nm sample the non-magnetic plasma  and non-magnetic Drude models are excluded by the data at the 95\% confidence level for separations $z<350$~nm. The  $t= 84$~nm sample has very small signal consistent with all remaining  models. The  interaction between the $t= 21$~nm and the Ni-coated sphere shows that the non-magnetic models (plasma and Drude) can be rejected at 68\% confidence level. However, all three models (except magnetic Drude) coincide with the data at the 95\% confidence level.

\acknowledgments

This work was performed, in part, at the Center for Nanoscale Materials, a U.S. Department of Energy, Office of Science, Office of Basic Energy Sciences User Facility under Contract No. DE-AC02-06CH11357.  R.S.D. acknowledges financial and technical support from the IUPUI Nanoscale Imaging Center, the IUPUI Integrated Nanosystems Development Institute, and the Indiana University Center for Space Symmetries. 

\section*{APPENDICES}

\subsection{Force correction due to sample topography.}

In this Appendix  Eq. (\ref{heicor}) is demonstrated. Since the spatial scale over which the height profile $h(r,\phi)$ changes appreciably is much larger than the Casimir interaction radius $\rho$, the separation $z$ can be considered as locally constant.  Therefore, the force $F_h$ taking account of the surface topography can be estimated by replacing in Eq. (\ref{meas}) $z$ by the local separation $z-h(r, \phi)$:
\be
F_h= -\frac{{\rm i}}{2} \int_{0}^{ 2 \pi} d \phi \; F_{\rm C}(z-h(\phi); \phi)\;e^ { {\rm i}\,n \phi}\;,\label{meas1}
\ee
where for brevity  the constant radius $r$ is omitted.
To estimate $F_h$  the step-wise approximation of the Casimir force Eq. (\ref{step}) is used and
$$
F_h= -\frac{{\rm i}}{2} \int_{0}^{ 2 \pi} d \phi \; [F_{\rm Au}(z-h(\phi)) \chi(n \phi)
$$
\be
+\; F_{\rm Ni}(z-h(\phi)) \chi(n \phi-\pi)]  \;e^ { {\rm i}\,n \phi}\;
\ee
is found. Using the identity $\chi(\theta-\pi)=1-\chi(\theta)$, the above expression can be recast as 
\be
F_h= -\frac{{\rm i}}{2} \int_{0}^{ 2 \pi} d \phi \; [F_{\rm Ni}(z-h(\phi))+F(z-h(\phi))  \chi(n \phi) ]  \;e^ { {\rm i}\,n \phi}\;.\label{hei1}
\ee
Since $|h(\phi)| \ll z$, a Taylor expansion of the above expression in powers of the height profile can be done.  Upon expanding the first term between the square brackets on the r.h.s. of Eq. (\ref{hei1}) to first order in $h$ and the second one to second order in $h$, the force  correction $\delta F$ 
$$
\delta F(z) = -\frac{3 \, {\rm i}}{2}F_{\rm Ni}(z)  \, \frac{h_n}{z}  
$$
\be
- \frac{3 \,{\rm i}}{2}F(z) \int_{0}^{ 2 \pi} d \phi \, \left[\frac{h(\phi)}{z}+ 2 \frac{h^2(\phi)}{z^2}   \right] \chi (n \phi) \, e^{ {\rm i}\,n \phi}\;
\ee
is found, where $h_n=\int_0^{2 \pi} d \phi \,h(\phi) \exp({\rm i} n \phi)$ is the $n$-th Fourier coefficient of $h(\phi)$. In writing the above expression it was used that both $F_{\rm Ni}(z)$ and $F(z)=F_{\rm Au}(z)-F_{\rm Ni}(z)$  have an approximate power-like behavior $F_{\rm Ni}(z) \sim z^{-\alpha_{\rm Ni}}$, $F(z) \sim z^{-\alpha}$, with  exponents $\alpha$ and $\alpha_{\rm Ni}$ both close to 3. Finally, by substituting into the above formula the Fourier transform 
\be
\chi(\theta)=\frac{1}{2}-\frac{{\rm i}}{\pi} \sum_{p=0}^{\infty} \frac{e^{{\rm i} (2 p -1) \theta}}{2 p -1}\;,
\ee 
 it is found that
$$
\delta F(z) = -\frac{3 \, {\rm i}}{4}(F_{\rm Au}(z) +F_{\rm Ni}(z)) \, \frac{h_n}{z}  
$$
\be
- \frac{3}{2 \, \pi}F(z)  \, \sum_{p=0}^{\infty}  \frac{1}{2 p -1}\left[\frac{h_{2 p n}}{z}+ 2 \frac{(h^2)_{2  p n }}{z^2}   \right]  \;,
\ee
where $(h^2)_m$ denotes the $m$-th Fourier coefficient of $h^2(\phi)$. The summands with $p \neq 0$ between the square brackets involve  Fourier coefficients  of the height profile and its square of order $2 p n$, that are multiples of the large number $n=300$. Since the height profile $h(\phi)$ is a slowly varying function, all these summands can be neglected compared to the $p=0$ summand. Bearing in mind that $h_0=0$
\be
\delta F(z) =  -(F_{\rm Au}(z) +F_{\rm Ni}(z)) \, \frac{3 \, {\rm i} h_n}{4 z}  + 6\, F(z)  \,  \frac{\langle h^2 \rangle  }{z^2}    \;,
\ee
where $\langle h^2 \rangle$ is the angular average of $h^2(\phi)$.

\subsection{The Casimir force  near Au-Ni boundaries\label{two}}

In this Appendix  the Casimir force $F_{\rm C}(z;r,\phi)$ at points close to the Au-Ni boundaries is estimated.  To be definite, suppose that the sphere tip is approaching  the Ni-Au boundary placed at $\phi=0$. Since the width $L=78.5\,\mu$m of the Au and Ni sectors is much larger than $\rho$, it is safe to assume that the  boundaries placed at $\phi=\pm  \pi/n$ are infinitely far away. Therefore, to study the transition region near $\phi=0$ no significant error is made  thinking that the Ni region to the left and the Au region to the right extend all the way to infinity.  Consider a Cartesian coordinate system $(x,y)$ in the plane $S$  containing the sample surface, such that the  boundary $\phi=0$ coincides with the $y$ axis. The $x$ axis is oriented such that the half-plane $S_+$ with $x>0$ corresponds to the Au sector, while the half-plane $S_-$ with $x<0$ corresponds to the Ni sector. Suppose that the sphere tip of a large sphere ($R \gg z$) is above the point $P$ of $S$ of coordinates $(s,0)$ along the $x$ axis.    To estimate the Casimir force $F_{\rm C}(z,s)$ on the sphere the simple  Derjaguin additive approximation is used.\cite{derj} This approximation  expresses $F_{\rm C}(z,s)$ as the sum of the elementary Casimir forces  on the surface elements of the sphere, regarded as small portions of a plane interacting only with the surface elements of $S$ that lie right under them. Within this approximation, the surface elements of the sphere above the   half-plane $S_-$  only see Ni, while those above the half-plane $S_+$   only see Au. Representing with $ F^{(\rm pp)}_{\rm Ni}(z)$ and $ F^{(\rm pp)}_{\rm Ni}(z)$  the respective  Casimir pressures for two parallel plates at distance $z$,  $F_{\rm C}$ is expressed as 
$$
F_{\rm C}(z,s)=\int_{-\infty}^0 dx \int_{-\infty}^{\infty} dy\; F_{\rm Ni}^{(\rm pp)}(d(x,y;z,s))
$$
\be
+ \int_0^{\infty} dx \int_{-\infty}^{\infty} dy \;F_{\rm Au}^{(\rm pp)}(d(x,y;z,s))\;,
\ee   
where   $d(x,y;z,s)=z+R-\sqrt{R^2-(x-s)^2-y^2}$ is the  height    of the sphere surface element $dx dy$  whose center  is above the point of $S$ of coordinates $(x,y)$. To compute the integral  a polar coordinate system $r,\theta$ in $S$, with origin at $P$, is used. Expressing $ F_{\rm Ni/Au}^{(\rm pp)}$ in terms of the respective Casimir free energies $F_{\rm Ni/Au}^{(\rm pp)}(z)=-d {\cal F}_{\rm Ni/Au}/dz$
\be
F_{\rm C}(z,s)= -\int \int_{S_-}  d \theta d r\,r   \;  \frac{\partial {\cal F}_{\rm Ni}}{\partial z}
-\int \int_{S_+}  d \theta d r\,r   \;    \frac{\partial {\cal F}_{\rm Au}}{\partial z}\;.
\ee   
The integral over $r$ can be done easily,   if one observes  that  the points of the sphere which contribute significantly are close to the tip. For those points $d \simeq z+ r^2/2 R$, and then
${\partial {\cal F}_{\rm Au/Ni}}/{\partial z}  \simeq 2 R  {\partial {\cal F}_{\rm Au/Ni}}/{\partial (r)^2} $.  After   the change of variables $\theta\rightarrow y=|s| \,\tan\theta$ in the final integral over $\theta$,  it is then easy to obtain the following formula valid for $s<0$:
\begin{widetext}
\be
F_{\rm C}(z,s)= F_{\rm Ni}(z)+ \frac{1}{2 \pi}\int_{-\infty}^{\infty} \frac{|s|\,dy}{s^2+y^2} \left[F_{\rm Au}(z+(s^2+y^2)/2R)- F_{\rm Ni}(z+(s^2+y^2)/2R)\right]\;.\label{derj}
\ee
\end{widetext}
For $s>0$, one has to exchange $F_{\rm Ni}(z)$ by $F_{\rm Au}(z)$. When Eq. (\ref{derj}) is evaluated numerically, it is found that $F_{\rm C}(z,s)$ becomes quickly indistinguishable from
$ F_{\rm Ni}(z)$  ($ F_{\rm Au}(z)$)  for $s< - \rho$ ($s > \rho$). For $|s| < \rho$,  $F_{\rm C}(z,s)$ is well approximated by a straight line  $F_{\rm C}(z,s)\approx (F_{\rm Au}(z)+F_{\rm Ni}(z))/2+(F_{\rm Au}(z)-F_{\rm Ni}(z)) s/2 \rho$. 

\subsection{Errors}

A table of possible errors in the measurements and their interpretation are listed in table~\ref{tableI}. Unless otherwise noted they correspond to a separation of $z = 200$~nm and for $\tau = 3000$~s. The effect of experimental uncertainties (thicknesses, optical properties) on the theoretical calculations are not included since they were extensively discussed and plotted in Section~\ref{comp}.

\begin{table} [htbp]
	\centering
	\begin{tabular}{ |p{2.1cm}||p{3.0cm}|p{1cm}| p{2cm}| }
		\hline
		Origin &  $\delta F$ & Sample & Method\\
		\hline
		$\Delta z$   &   $< (10^{-4} F + 10^{-6} F_{\rm C})$ &   All & Calculation\\
		Patches    & $<$ 0.03 fN   & All & Ref.~[\onlinecite{Patches}]\\
		Mag. domains & $<$ 0.05 fN &  Ni  & Measurement\\
		Size effects &  $< 10^{-3} F$    & All & Calculation \\
		Spike&  $<$ 0.01 fN & All & Measurement\\
		Wobble&  $<$ 0.2 fN & All & Measurement\\
		$\Delta R$ & $< 10^{-3} F$ & All & Calculation\\
		$\Delta r$&  $< 10^{-4} F $ & All & Calculation\\
		PFA &   $< 1.5 \times 10^{-3} F$ & All & Calculation\\

		\hline 
	\end{tabular}
	\caption{Systematic contributions to the measured force. Columns show the origin of the systematic effect, its effect, samples over which the systematic effect plays a role (i.e. Ni-covered sphere or both), and the methods for obtaining $\delta F$, respectively.  $F$ is the force measured in this experiment, $F_{\rm C}$ represents the static Casimir force between the sphere and the sample.}
	\label{tableI}
\end{table}


\section*{References}

 



\begin{thebibliography}{200}

\bibitem{vdW} V. A. Parsegian, {\it van der Waals Forces: A Handbook for Biologists, Chemists, Engineers, and Physicists} (Cambridge University Press, Cambridge, 2005).

\bibitem{Casimir} H. G. B. Casimir, Proc. Nederl. Akad. Wetensch. {\bf 51}, 793 (1948).

\bibitem{RSDLamoreaux1997}S. K.~Lamoreaux,  Phys. Rev. Lett. {\bf 78}, 5 (1997).

\bibitem{RSDMohideen1998}U.~ Mohideen and A.~Roy,  Phys. Rev. Lett. {\bf 81}, 4549 (1998).

\bibitem{RSDEderth2000}T.~Ederth,   Phys. Rev. A {\bf 62}, 062104 (2000).

\bibitem{RSDBressi2001}G.~Bressi, G.~Carugno, R.~Onofrio, and G.~Ruoso,  Phys. Rev. Lett. {\bf 88},  041804 (2002).

\bibitem{Brown2005} M.~Brown-Hayes, D. A. R.~Dalvit, F. D.~Mazzitelli, W. J.~Kim, and R.~Onofrio, Phys. Rev. A {\bf 72}, 052102 (2005). 

\bibitem{RSDChan2001a}H. B.~Chan, V. A.~Aksyuk, R. N.~Kleiman, D. J.~Bishop, and F.~Capasso,   Science {\bf 291}, 1941 (2001);   Phys. Rev. Lett. {\bf 87}, 211801 (2001).

\bibitem{RSDTang2012}D. Garc\'{\i}a-S\'anchez, K. Y. Fong, H. Bhaskaran, S. Lamoreaux, and H. X. Tang,  Phys. Rev. Lett. {\bf  109}, 027202 (2012).

\bibitem{RSDChan2013}J.~Zou, Z.~Marcet, A. W.~Rodriguez, M. T. H.~Reid, A.P.~McCauley, I. I.~Kravchenko, T.~Lu, Y.~Bao, S. G.~Johnson, and H. B.~Chan,  Nat. Commun. {\bf 4}, 1845 (2013).

\bibitem{RSDDecca2003}R.S. Decca, D. L\'opez, E. Fischbach and D.E. Krause,  Phys. Rev. Lett. {\bf 91}, 050402 (2003).

\bibitem{RSDIanuzzi2004}D. Iannuzzi, M. Lisanti, and F. Capasso,  Proc. Nat. Acad. Sci. {\bf 101}, 4019 (2004).

\bibitem{RSDMohideen2005}F. Chen, U. Mohideen, G. L. Klimchitskaya, and V. M. Mostepanenko, Phys. Rev. A {\bf 72}, 020101 (2005).

\bibitem{RSDLamoreaux2009}W. J. Kim, A.O. Sushkov, D. A. R. Dalvit, and S. K. Lamoreaux,  Phys. Rev. Lett. {\bf 103}, 060401 (2009).

\bibitem{RSDMunday2009}J. N. Munday, F. Capasso, and V. A. Parsegian,  Nature {\bf 457}, 170 (2009).

\bibitem{deMan} S.~de Man, K.~Heeck, R. J.~Wijngaarden, and D.~Iannuzzi, Phys. Rev. Lett. {\bf 103}, 040402 (2009).

\bibitem{RSDPalasantzas2010}G. Torricelli, P. J. van Zwol, O. Shpak, C. Binns, G. Palasantzas, B. J. Kooi, V. B. Svetovoy, and M. Wuttig,  Phys. Rev. A {\bf 82}, 010101 (2010).

\bibitem{RSDMohideen2012}A. A. Banishev, C.-C. Chang, R. Castillo-Garza, G.  L. Klimchitskaya, V. M. Mostepanenko, and U. Mohideen,  Phys. Rev. B {\bf 85}, 045436 (2012).

\bibitem{RSDDecca2009}R.S.Decca, D. L\'opez, and E. Osquiguil,  Int. J. Mod. Phys. A {\bf 25}, 2223 (2010).

\bibitem{RSDMohideen2013b}R.~Castillo-Garza, J.~Xu, G. L.~Klimchitskaya, V. M.~Mostepanenko, and U.~Mohideen,  Phys. Rev. B {\bf 88}, 075402 (2013).

\bibitem{book2} M.~Bordag, G. L.~Klimchitskaya, U.~Mohideen, and V. M.~Mostepanenko,  {\it Advances in the Casimir Effect} (Oxford University Press, Oxford, 2009).

\bibitem{reviewCasimir} See, for example,   G. L. Klimchitskaya, U. Mohideen, and V. M. Mostepanenko, Rev. Mod. Phys. {\bf 81}, 1827 (2009).

\bibitem{capasso} A. W.~Rodriguez, F.~Capasso, and S. G.~Johnson, Nature Photon. {\bf 5}, 211 (2011).

\bibitem{decca4} R. S. Decca, D. L\'{o}pez, E. Fischbach,  G. L. Klimchitskaya, D. E. Krause, and V. M. Mostepanenko, Annals Phys. {\bf 318}, 37 (2005).

\bibitem{Lifshitz} E. M. Lifshitz, Zh. Eksp. Teor. Fiz. {\bf 29}, 94 (1955) [Sov. Phys. JETP {\bf 2}, 73 (1956)].


\bibitem{DeccaEJP2007} R. S. Decca, D. L\'opez, E. Fishbach, G. L. Klimchitskaya, D. E. Krause, and V. M. Mostepanenko, Eur. Phys. J. C {\bf 51}, 963 (2007).

\bibitem{MohideenPRA2012} A. A. Banishev, C.-C. Chang, G. L. Klimchitskaya, V. M. Mostepanenko, and U. Mohideen, Phys. Rev. B {\bf 85}, 195422 (2012);  A. A. Banishev, G. L. Klimchitskaya, V. M. Mostepanenko, and U. Mohideen, Phys. Rev. B {\bf 88}, 155410 (2013); Phys. Rev. Lett. {\bf 110}, 137401 (2013).

\bibitem{LamoreauxNat2011} A. O. Sushkov, W. J. Kim, D. A. R. Dalvit, and S. K. Lamoreaux, Nature Phys. {\bf 7}, 230 (2011).

\bibitem{BimontePRL2014} G. Bimonte, Phys. Rev. Lett. {\bf 112}, 240401 (2014); Phys. Rev. Lett. {\bf 113}, 240405 (2014); Phys. Rev. B {\bf 91}, 205443 (2015); Phys. Rev. A {\bf 92}, 032116 (2015).

\bibitem{DeccaPRL2014} Y.-J. Chen, W. K. Tham, D. E. Krause, D. L\'opez, E. Fischbach, and R. S. Decca, {\it Isoelectronic measurements yield stronger limits on hypothetical Yukawa interactions in the 40--8000 nm range},  arXiv:1410.7267v1.

\bibitem{patches}R. O. Behunin, F. Intravaia, D. A. R. Dalvit, P. A. Maia Neto, and S. Reynaud, Phys. Rev. A {\bf 85}, 012504 (2012).

\bibitem{Kolb} P. W. Kolb, R. S. Decca, and H. D. Drew. Rev. Sci. Instrum.{\bf 69}, 310-312 (1998).

\bibitem{Casimirbook} W. Simpson and U. Leonhardt Eds.,  {\it Forces of the quantum vacuum: An introduction to Casimir physics} (World Scientific, Singapore, 2015), chapter 4.

\bibitem{elcal} R. S. Decca and D. L\'opez, Int. J. Mod. Phys. A {\bf 24}, 1748 (2009).

\bibitem{fit} F. Chen, U. Mohideen, G. L. Klimchitskaya and V. M. Mostepanenko, Phys. Rev. A {\bf 74}, 022103 (2006).

\bibitem{Casimirz}  A measurement of the Casimir force when the sphere is over the interstitial sector also yields the same value of $z$ as the WLI or the electrostatic measurement. This procedure, however, assumes the Casimir force is known. Consequently it was not used in the paper.

\bibitem{StPet} R. S. Decca, Int. J. Mod. Phys. A {\bf 31}, 1641024 (2016).

\bibitem{footnote} The force $F$ has been measured using either a Ni or a Au coated sphere. When necessary,  a superscript enclosed in parenthesis   denote the material of the sphere's coating. Thus, for example, $F^{(\rm Au)}_{\rm Ni}(z)$ shall denote the force $F_{\rm Ni}(z)$  on a Au-coated sphere. 

\bibitem{richmond} P. Richmond and B. W. Ninham, J. Phys. C {\bf 4}, 1988 (1971).

\bibitem{tomas} M. S. Toma\u{s}, Phys. Lett. A {\bf 342}, 381 (2005).

\bibitem{born} M. Born and E. Wolf, {\it Prinicples of Optics} (Pergamon Press, Oxford) 1959.

\bibitem{Yeh} P. Yeh, {\it Optical Waves in Layered Media} (John Wiley \& Sons, New York) 1988.

\bibitem{palik}  {\it Handbook of Optical Constants of Solids}, edited by E. D. Palik (Academic, New York, 1995).


\bibitem{generKK} G. Bimonte, Phys. Rev A {\bf 81}, 062501 (2010).

\bibitem{generKK2} G. Bimonte, Phys. Rev A  {\bf 83}, 042109 (2011).



\bibitem{ordal} M.A. Ordal, R.J. Bell, R.W. Alexander, L.L. Long, and M. R. Querry, App. Opt. {\bf 24}, 4493 (1985).

\bibitem{fosco2}  C. D. Fosco, F. C. Lombardo, and F. D. Mazzitelli, Phys. Rev.D {\bf 84}, 105031 (2011).

\bibitem{grad1} G. Bimonte, T. Emig, R. L. Jaffe, and M. Kardar, EPL {\bf 97}, 50001 (2012).

\bibitem{grad2} G. Bimonte, T. Emig, and M. Kardar, Appl. Phys. Lett. {\bf 100}, 074110 (2012).

\bibitem{teo} L. P. Teo, Phys. Rev. D {\bf 88}, 045019 (2013).

\bibitem{raoul} R. Esquivel-Sirvent and V. B. Svetovoy, Phys. Rev. B {\bf 72}, 045443 (2005).

\bibitem{volokitin} A.I. Volokitin and B.N.J. Persson, Phys. Rev. B {\bf 78}, 155437 (2008); {\it ibid.} {\bf 81}, 239901 (2010). 

\bibitem{lucy} S. Lucyszyn, PIERS Online {\bf 4}, 686 (2008).

\bibitem{derj} B.V. Derjaguin, Kolloid. Z. {\bf 69}, 155 (1934).

\bibitem{Patches}  R. O. Behunin, D. A. R. Dalvit, R. S. Decca, and C. C. Speake, Phys. Rev. D {\bf 89}, 051301(R) (2014); R. O. Behunin, D. A. R. Dalvit, R. S. Decca, C. Genet, I. W. Jung, A. Lambrecht, A. Liscio, D. L\'opez, S. Reynaud, G. Schnoering, G. Voisin, and Y. Zeng, Phys. Rev. A {\bf 90}, 062115 (2014). 

\bibitem{break} An experimental accident precluded the measurement between the $t=21$~nm sample and the Au-coated sphere. 

\end{thebibliography}
\end{document}